\documentclass[journal,lettersize]{IEEEtran}

\pdfpageattr{/Group << /S /Transparency /I true /CS /DeviceRGB >>}

\usepackage[utf8]{inputenc}
\usepackage[T1]{fontenc}
\usepackage{textcomp}
\usepackage{stfloats}
\usepackage{fixltx2e}
\usepackage{microtype}
\usepackage{calc}
\usepackage[normalem]{ulem}
\usepackage{eurosym}
\usepackage[bottom]{footmisc}

\usepackage[range-phrase=--,per-mode=symbol-or-fraction,binary-units=true,range-units=single,list-units=single,detect-all,forbid-literal-units]{siunitx}
\usepackage{silence}
\AtBeginDocument{
\DeclareSIUnit\kmh{\km\per\hour}
\DeclareSIUnit\mps{\m\per\s}
\DeclareSIUnit\vehicle{veh}
\DeclareSIUnit\lane{lane}
\DeclareSIUnit\flow{\vehicle\per\hour}
\DeclareSIUnit[per-mode=repeated-symbol]{\density}{\vehicle\per\km\per\lane}
\DeclareSIUnit[per-mode=repeated-symbol]{\lanecapacity}{\vehicle\per\lane\per\hour}
\DeclareSIUnit\ghz{\giga\hertz}
\DeclareSIUnit\hz{\hertz}
\DeclareSIUnit\km{\kilo\metre}
\DeclareSIUnit\m{\metre}
\DeclareSIUnit\s{\second}
\DeclareSIUnit\eu{\text{\euro{}}}
}

\usepackage{csquotes}
\usepackage[backend=bibtex,style=ieee,citestyle=numeric-comp,doi=false,isbn=false,mincitenames=1,maxcitenames=2,minbibnames=1,maxbibnames=99]{biblatex}
\DeclareFieldFormat{sentencecase}{#1} 
\DeclareFieldFormat{titlecase}{#1} 
\usepackage{xpatch}
\xpatchbibmacro{textcite}{\addspace}{\addnbspace}{}{}
\xpatchbibmacro{Textcite}{\addspace}{\addnbspace}{}{}
\DefineBibliographyStrings{english}{
    andothers = et~al\adddot\addspace
}

\usepackage{amsmath}

\usepackage{amssymb}
\usepackage{amsfonts}
\usepackage{amsthm}

\usepackage{booktabs}
\usepackage{url}

\usepackage[inline]{enumitem}

\usepackage[caption=false,font=footnotesize]{subfig}
\usepackage[pdftex]{graphicx}
\usepackage{pgfplots}
\DeclareGraphicsExtensions{.pdf,.png,.jpg}

\usepackage{todonotes}

\usepackage{tikz}
\usetikzlibrary{arrows}
\usetikzlibrary{arrows.meta}
\usetikzlibrary{calc}
\usetikzlibrary{chains}
\usetikzlibrary{scopes}
\usetikzlibrary{shapes}

\usepackage[american]{babel}
\usepackage{hyphenat}

\usepackage[noend]{algorithmic}
\usepackage{algorithm}

\usepackage{hyperref}
\usepackage{hypcap}

\usepackage[capitalize,noabbrev]{cleveref}
\crefname{paragraph}{Paragraph}{Paragraphs}
\Crefname{paragraph}{Paragraph}{Paragraphs}
\labelformat{subsection}{\thesection-#1}

\usepackage[nodebug]{flushend}

\RequirePackage{xstring}
\RequirePackage{xparse}
\RequirePackage[]{acro}
\NewDocumentCommand\acrodef{mO{#1}mG{}}{\DeclareAcronym{#1}{short={#2}, long={#3}, #4}}
\acrodef{ACC}{adaptive cruise control}
\acrodef{CACC}{cooperative adaptive cruise control}{alt={cooperative adaptive cruise controller}}
\acrodef{C-V2X}{cellular \acl*{V2X}}
\acrodef{DSRC}{distributed short-range communication}
\acrodef{EV}{electric vehicle}
\acrodef{HBEFA}{handbook of emission factors for road transport}
\acrodef{ITS}{intelligent transportation system}{short-plural-form={ITS}}
\acrodef{V2X}{vehicle-to-everything}
\acrodef{VLC}{visible light communication}
\acrodef{V-VLC}{vehicular \acl*{VLC}}

\usepackage[color]{changebar}
\def\todo{%
    \cbcolor{magenta}
    \cbstart%
    \begingroup
    \color{magenta}
    \obeylines%
    \begingroup\lccode`~=`\^^M\lowercase{\endgroup\def~}{\par\leavevmode}%
    \parindent0em%
    \catcode`\_=\active
    \catcode`\<=\active\lccode`~=`<\lowercase{\def~}{$<$}%
    \catcode`\>=\active\lccode`~=`>\lowercase{\def~}{$>$}%
    \catcode`\#=\active\lccode`~=`\#\lowercase{\def~}{$\#$}%
    \catcode`\^=\active\lccode`~=`\^\lowercase{\def~}{$\hat{~}$}%
    \todoCtd
}\def\todoCtd#1{%
    TODO: #1%
    \ifx&#1&...\fi%
    \endgroup
    \cbend
    \relax
}
\usepackage{comment}
\usepackage{draftfigure}

\NewDocumentCommand\IEEE{ s m d[] }{%
    \IfBooleanTF{#1}{}{IEEE\,}
    \nolinebreak[2]
    #2%
    \IfNoValueTF{#3}{%
    }{%
        \StrGobbleLeft{#3}{1}[\sommerIEEEFirstLetter]%
        \IfEq{\sommerIEEEFirstLetter}{}{%
            #3
        }{%
            \nolinebreak[3]
            \StrLeft{#3}{1}%
            \sommerIEEELettersSlashed{\sommerIEEEFirstLetter}%
        }%
    }%
}
\newcommand{\sommerIEEELettersSlashed}[1]{%
    /
    \StrLeft{#1}{1}%
    \StrGobbleLeft{#1}{1}[\sommerIEEESubsequentLetter]%
    \IfEq{\sommerIEEESubsequentLetter}{}{%
    }{%
        \sommerIEEELettersSlashed{\sommerIEEESubsequentLetter}
    }%
}

\newenvironment{notes}{%
    \textcolor{red}{Julian's notes:}
    \color{teal}
    \bgroup\obeylines%
}{\egroup}

\newcommand{\simulator}{PlaFoSim}
\newcommand{\p}{\IEEE{802.11}[p]}

\newcommand{\vtp}{vehicle-to-platoon}
\newcommand{\human}{\emph{human driving}}
\newcommand{\acc}{\emph{\acs*{ACC}}}
\newcommand{\similaritybased}{\emph{similarity-based platooning}}
\newcommand{\tripcostbased}{\emph{trip cost-based platooning}}

\usepackage{orcidlink}

\begin{document}

\title{Incentive-based Platoon Formation: Optimizing the Personal Benefit for Drivers}
\markboth{Incentive-based Platoon Formation: Optimizing the Personal Benefit for Drivers}{Heinovski \textit{et al.}}

\author{%
\IEEEauthorblockN{%
Julian Heinovski%
\textsuperscript{\orcidlink{0000-0003-3169-8109}}%
\IEEEauthorrefmark{1},%
\IEEEmembership{~Graduate~Student~Member,~IEEE}%
,%
Do\u{g}analp Ergen\c{c}%
\textsuperscript{\orcidlink{0000-0003-4640-031X}}%
\IEEEauthorrefmark{1},%
\IEEEmembership{~Member,~IEEE}%
,%
~%
\\
Kirsten Thommes%
\textsuperscript{\orcidlink{0000-0002-8057-7162}}%
\IEEEauthorrefmark{2}%
,%
~%
and
Falko Dressler%
\textsuperscript{\orcidlink{0000-0002-1989-1750}}%
\IEEEauthorrefmark{1},%
\IEEEmembership{~Fellow,~IEEE}%
}%
\thanks{%
Julian Heinovski, Do\u{g}analp Ergen\c{c}, and Falko Dressler are with the School of Electrical Engineering and Computer Science, TU Berlin, Germany;
E-Mails: \{heinovski,ergenc,dressler\}@ccs-labs.org;
Kirsten Thommes is with the School of Business Administration and Economics, Paderborn University, Germany;
E-Mail: kirsten.thommes@uni-paderborn.de.
}%
}

\maketitle

\begin{abstract}\nohyphens{%
Platooning or \ac{CACC} has been investigated for decades, but debate about its lasting impact is still ongoing.
While the benefits of platooning and the formation of platoons are well understood for trucks, they are less clear for passenger cars, which have a higher heterogeneity in trips and drivers' preferences.
Most importantly, it remains unclear how to form platoons of passenger cars in order to optimize the personal benefit for the individual driver.
To this end, in this paper, we propose a novel platoon formation algorithm that optimizes the personal benefit for drivers of individual passenger cars.
For computing \vtp{} assignments, the algorithm utilizes a new metric that we propose to evaluate the personal benefits of various driving systems, including platooning.
By combining fuel and travel time costs into a single monetary value, drivers can estimate overall trip costs according to a personal monetary value for time spent.
This provides an intuitive way for drivers to understand and compare the benefits of driving systems like human driving, \ac{ACC}, and, of course, platooning.
Unlike previous similarity-based methods, our proposed algorithm forms platoons only when beneficial for the driver, rather than solely for platooning.
We demonstrate the new metric for the total trip cost in a numerical analysis and explain its interpretation.
Results of a large-scale simulation study demonstrate that our proposed platoon formation algorithm outperforms normal \acs{ACC} as well as previous similarity-based platooning approaches by balancing fuel savings and travel time, independent of traffic and drivers' time cost.
}\end{abstract}


\acresetall%
\IEEEpeerreviewmaketitle%

%

\section{Introduction}%
\label{sec_introduction}



\IEEEPARstart{R}{oad} traffic has consistently grown in recent years, leading to increasing congestion and environmental pollution.
To cope with these adverse effects, modern vehicles are being equipped with (advanced) driver assistance systems and \acs{V2X} communication technologies like 5G-based \ac{C-V2X} and \p{}-based \acs{DSRC}, which allow vehicles to cooperate.
These technologies improve not only driving safety and comfort but also efficiency, enabling new \ac{ITS} solutions like \ac{CACC} and platooning~\cite{dressler2019cooperative,locigno2022cooperative}.
Today, classic \ac{ACC} is the \emph{de facto} standard for all new cars as well as for (semi-)automated driving on the freeway.
Going one step further, vehicular platooning allows multiple vehicles to drive in convoys with small but stable safety gaps using \ac{CACC}.
While both \ac{ACC} and \ac{CACC} improve traffic flows and safety, platooning additionally increases road utilization and reduces fuel consumption due to the slipstream effect~\cite{pi2023automotive,locigno2022cooperative,sturm2020taxonomy}.
The benefits of \ac{ACC} and platooning have been researched for decades but debate about the lasting impact of platooning is still ongoing~\cite{martinez-diaz2024impacts,braiteh2024platooning}.

String-stable operation of platoons has been demonstrated~\cite{%
shladover1991automated, 
rajamani2000demonstration, 
horowitz2000control, 
jootel2012final, 
jia2016survey,
locigno2022cooperative,
}
and cooperative join maneuvers have been designed~\cite{%
segata2014supporting,
maiti2019impact,
santini2019platooning,
paranjothi2020pmcd,
lee2021decentralized,
liu2022optimizing,
li2022review,
scholte2025experimental,
}
to form effective platoons.
The integration of human driven and automated (platooning) cars is a known issue~\cite{dressler2018cpss}.
It has been addressed in several works, mainly on the control side.
For example, first model-predictive control solutions take human reactions into account~\cite{kennedy2023centralized}.
However, deciding which vehicles best form a platoon together remains an unresolved challenge towards large-scale deployment of platooning~\cite{hou2023large-scale}.

Although simple ad-hoc approaches facilitate the rapid setup of platooning~\cite{maiti2019analysis,ghiro2024platoons}, they typically rely solely on the current position of vehicles and assume platooning is universally desired.
Computing \vtp{} assignments, i.e., assigning vehicles to platoons, requires more complex computations, typically aiming at optimizing macroscopic objectives such as a smoother traffic flow and better road utilization.
However, there may be a trade-off between these macroscopic results and individualized optimization as time and cost savings for the individual vehicle may not be possible for everybody.

Platooning is primarily promoted for trucks due to its significant fuel-saving potential~\cite{locigno2022cooperative,martinez-diaz2024impacts,braiteh2024platooning}.
Trucks typically perform long trips, allowing them to share platoon benefits for a long time.
In contrast, passenger cars typically undertake more spontaneous, varied, and less predictable trips, making the benefits of platooning less evident compared to the recurring and structured routes of trucks in freight transport.
Given the resulting complexity, they are still a less interesting use case for platooning from the perspective of the automotive industry.
Additionally, so far, individual drivers of passenger cars cannot precisely assess their immediate benefits due to complex and individualized utility functions consisting of components difficult to offset, e.g., travel time, fuel savings, and increased safety and comfort.
These depend on the unique expectation of the individual driver~\cite{heinovski2024platooning} and personal preferences and values, so that the drivers' individual and estimated benefits should be considered~\cite{dai2015personalized,lesch2021overview,heinovski2024where}.

A natural and thus intuitive incentive for drivers to employ platooning is a potential reduction in the overall trip cost (cf.\ \cite{niyato2016economics}).
For instance, an optimal driving speed can result in less fuel cost but causes an extended trip time compared to driving faster yet fuel-inefficient.
Vice versa, a shorter travel time often leads to sub-optimal emissions and fuel cost~\cite{sommer2010emissions}.
Unlike personal preferences, fuel costs and travel time can be captured in a single monetary value:
While the fuel cost is dependent on the vehicle and the driving speed, the opportunity cost for travel time and the desired level of fuel-efficient driving depends on the individual driver and their personal preferences (cf.\ \cite{dai2015personalized}).
In general, travel time can be measured as an opportunity cost~\cite{kouwenhoven2018value,small2012valuation}.
Opportunity costs are the costs one has to ``pay'' because they are traveling and cannot perform productive tasks during the driving time.
Thus, there is always a trade-off between fuel consumption as a variable cost of driving and time consumption as opportunity cost of travel time.
These are the two important key metrics that the driver has to balance on to employ and also assess the individual driving systems~\cite{sommer2010emissions,sun2016save}.

Accordingly, in this paper, we extend the understanding of personal benefit by proposing a metric that combines multiple optimization factors in one monetary unit.
In particular, we make use of the consumed fuel and the travel time as conceptually suggested in our earlier work~\cite{heinovski2024platooning}.
By assigning a personal monetary value for time spending, drivers can estimate their overall trip cost induced by different driving systems.
We also show how the personal monetization of time can be modeled by using real-world statistical data.

On top of this, we propose a novel platoon formation algorithm that utilizes the trip cost metric for computing \vtp{} assignments.
The algorithm optimizes drivers' personal benefits by comparing the cost of individual driving using \ac{ACC} with the cost of driving in a platoon, including the cost of joining and the expected benefits.
Unlike previous similarity-based methods, it forms platoons only when beneficial for the driver, rather than solely for platooning.

We perform a large-scale simulation study with \simulator{}~\cite{heinovski2021scalable} to compare our proposed algorithm against different driving systems including \ac{ACC} (as a baseline) as well as human driving and traditional platooning.
Our results show that standard \ac{ACC} always outperforms conventional human driving, but lacks behind platooning, especially in medium to high-density traffic.
Furthermore, our results demonstrate that our new platoon formation algorithm outperforms traditional similarity-based platooning in almost all situations, regardless of traffic density or drivers' time costs.

To the best of our knowledge, this is the first study that evaluates and optimizes the drivers' personal benefit from driving systems with a dedicated metric and a corresponding algorithm for platoon formation, respectively.

Our primary contributions can be summarized as follows:
\begin{itemize}
    \item We introduce and demonstrate a new metric to estimate the overall trip cost according to a personal monetary value for time spending.
    \item We developed a novel platoon formation algorithm that utilizes our proposed metric for computing \vtp{} assignments.
    \item In a large-scale simulation study, we show that our proposed algorithm can optimize drivers' personal benefit given as a monetary value.
    The benefit is maximized if there are many other drivers with the same monetary value for time spending.
\end{itemize}

The remainder of this paper is structured as follows:
We review related work from the literature in \cref{sec_related_work}.
We introduce our novel metric for the total trip cost and demonstrate its concept in a numerical analysis in \cref{sec_metric}.
We describe our novel algorithm for platoon formation utilizing our new metric in \cref{sec_platoon_formation}.
We illustrate our methodology for evaluating the proposed algorithm and report corresponding results in \cref{sec_evaluation}.
We discuss our evaluation results in \cref{sec_discussion} and conclude our findings in \cref{sec_conclusion}.

%

\section{Related Work}%
\label{sec_related_work}

\noindent 
We aim to form platoons of individual passenger cars that spontaneously start their trips on freeways.
As a result, their properties (e.g., destination) and drivers' preferences (e.g., desired speed) are unknown beforehand and can vary significantly.
Typical solutions designed for optimally forming platoons of trucks~\cite{bhoopalam2018planning}, where transport assignments and deadlines are predefined, are not applicable here.
While simple approaches enable ad-hoc platoon formation~\cite{maiti2019analysis,ghiro2024platoons}, they typically only consider vehicle positions and assume that platooning is always desired.
Since a vehicle's platoon choice significantly impacts the resulting benefits, its assignment should be optimized.
Therefore, assigning vehicles to platoons often requires more complex computations to optimize specific factors~\cite{sturm2020taxonomy}.
Various studies consider different optimization objectives and properties but often prioritize macro-level goals, such as a smoother traffic flow and better road utilization, rather than focusing on outcomes for individual vehicles~\cite{lesch2021overview}.

\subsection{Traditional Optimization Factors}%
\label{sec_related_work_traditional}

\noindent 
A natural optimization factor is fuel consumption, as platooning promises significant savings through the slipstream effect.
As a result, many studies aim to group platoons accordingly, analyzing how to efficiently execute catch-up and slow-down maneuvers~\cite{%
liang2013when,
larson2015distributed,
liang2016heavy-duty,
saeednia2016analysis,
wu2019energy-efficient,
maged2020behavioral,
bian2022fuel,
}.
However, most studies focus on trucks, which typically have a fleet speed limit of around \SIrange{80}{90}{\kmh} and exhibit minimal speed variance, making platooning more straightforward.
In contrast, passenger cars not only have higher technical speed capabilities but also exhibit a greater variance in speed preferences due to the heterogeneity of drivers, making platooning more challenging.
Additionally, these studies often do not consider other factors, such as route and travel time, assuming that platooning is always beneficial.

A common approach, often also used for passenger cars, is grouping vehicles based on their similarity in certain properties and driver preferences, often also combining multiple optimization factors.
By grouping vehicles with similar destinations or routes, the distance a platoon stays intact and vehicles can share platoon benefits is maximized~\cite{%
hall2005sorting, 
dao2013strategy, 
sharma2022potent, 
dokur2022edge, 
}.
This allows vehicles to save more fuel while increasing lane capacity and traffic throughput.
Additionally, studies propose grouping vehicles with similar (desired) driving speeds and positions~\cite{%
hoshino2013reactive, 
su2016autonomous, 
heinovski2024where, 
}.
Thereby, all platoon members have similar driving demands, resulting in less deviation from their preferences.
Additionally, join maneuvers are only performed with platoons in reasonable proximity.
Approaches for optimizing even more factors at the same time have been proposed, but the objectives \emph{``may or may not be compatible''} such that some way of further prioritization is required~\cite{willigen2013evolving}.
%
Although these approaches enable platooning benefits, grouping vehicles by similarity has several drawbacks:
It remains unclear which properties, or combinations thereof, yield the best results in terms of platooning benefits.
Furthermore, in some situations, not performing platooning and using only standard \ac{ACC} instead could be more beneficial~\cite{heinovski2024platooning}.

Other approaches partly resolve these drawbacks by increasing flexibility and also addressing the expected benefits with the use of a utility function~\cite{maiti2019analysis,malik2023unlocking}.
However, these utility functions are often abstract, and their significance is not immediately clear to drivers.
More generally, many approaches only consider the properties of the platooning opportunities for the assignment decision but ignore the join maneuver required for establishing the desired formation.

\subsection{Incentives and Cost}%
\label{sec_related_work_incentives}

\noindent 
\textcite{niyato2016economics} emphasize the importance of incentives such as cost, revenue, and profit for sustaining IoT development and operation, highlighting their relevance also for platooning.
A natural and intuitive incentive for drivers is the potential reduction in overall costs. 
%
%
%
In the context of \ac{ITS}, \textcite{sommer2010emissions} argue for balancing emissions and fuel consumption together with travel time, as optimizing for travel time alone often results in suboptimal emissions and fuel consumption.
This creates a fundamental trade-off between these key metrics that drivers must consider when assessing driving systems to employ~\cite{sommer2010emissions,sun2016save}.
While fuel cost depends on the vehicle and driving speed, opportunity cost for longer travel time varies based on individual factors,
such as how much one can earn when being productive instead of traveling (cf.\ \cite{dai2015personalized}).

Monetary rewards have been proposed as a direct incentive for promoting platooning.
\textcite{ledbetter2019lips} suggest rewarding drivers for taking on the role of platoon leaders, which is less fuel-efficient and more risky than being a follower.
Their payment system, based on estimated fuel savings and a leadership bonus, uses smart contracts to securely facilitate financial transactions in a distributed, trustless environment.
While this approach effectively prevents drivers from forfeiting leadership for personal gain, it does not address the process of forming the platoon.
%
Extending this idea, \textcite{earnhardt2022cooperative} propose a compensation scheme for trucks that perform catch-up and re-routing maneuvers to form platoons.
This incentivizes vehicles to sacrifice individual fuel economy for collective fuel savings.
However, this approach does not consider travel time, lacks personalization of driver preferences, and requires knowledge of vehicles' trajectories \& traffic patterns.

To address these limitations, \textcite{malik2023unlocking} use a multi-objective utility function including monetary rewards, such as insurance discounts and toll reductions, aiming for a mutually beneficial solution for authorities and drivers.
While their approach accounts for travel time, platoon distance, and speed differences, it does not fully consider the trade-offs between fuel savings and other factors.
Also, it assumes that platooning is always preferred and lacks personalization of drivers' benefits, which are unintuitively reported by the utility function.

\textcite{pelletier2021defining} analyze the costs of platoon formation by converting fuel and time costs into monetary equivalents, aiming for the optimal conditions to form platoons.
While the travel time is included in their cost metric, it is primarily used as a constraint in an optimization problem.
Their cost metric lacks personalization of drivers' benefits and its application is only briefly illustrated in a simple scenario involving two vehicles on a shared route, leaving more complex situations unexplored.

\subsection{Contribution}%
\label{sec_related_work_our_idea}

\noindent 
%
In earlier work~\cite{heinovski2024platooning}, we introduced a total trip cost metric that combines fuel consumption and travel time into a single monetary value for assessing different driving systems.
%
Beyond post-trip assessment, this metric can also be applied in real time.
By estimating the total cost before or during a trip, drivers can select the most beneficial driving system or platoon based on estimated savings in fuel and travel time.
This allows for a more intuitive decision-making process, offering practical insights into which option minimizes overall costs while being in line with drivers' personal preferences.
%
%
We aim to maximize drivers' individual benefits by considering their personal monetary value for time, offering a straightforward and intuitive decision-making tool.
Our approach ensures that platoons are formed only when they result in actual benefits for the driver rather than being enforced by default.

%

\section{The Personal Benefit of Driving Systems}%
\label{sec_metric}

\noindent 
In the following, we introduce our metric to quantify the drivers' personal benefit from (or incentive for) platooning.
Extending our preliminary results in~\cite{heinovski2024platooning}, we provide more elaborate explanations and a numerical analysis.

\subsection{Total Trip Cost Metric}%
\label{sec_metric_totaltripcost}

\noindent 
We aim for an intuitive metric that allows an easier trade-off between fuel consumption and travel time that drivers can easily grasp and use for understanding their personal benefit from driving systems, e.g., fuel and time savings due to reduced air drag and improved traffic flow in platooning~\cite{locigno2022cooperative}.
To achieve that, we propose to map both factors into a common monetary unit, allowing them to be combined to a single value for the overall cost of a trip.
In particular, we assess the total cost of a trip $C_{\text{trip}}$ by summing up the (estimated) cost of the consumed fuel and the (estimated) cost of the travel time as
\begin{equation}\label{eq_total_trip_cost_metric}%
C_{\text{trip}} = \text{fuel}_{\text{trip}} \cdot C_{\text{fuel}} + \text{time}_{\text{trip}} \cdot C_{\text{time}} \quad \text{.}
\end{equation}
%
Here, the cost of the consumed fuel is calculated as the product of the (estimated) fuel consumption for the trip ($\text{fuel}_{\text{trip}}$)%
\footnote{We illustrate the calculation and estimation of the fuel consumption in \cref{sec_metric_numerical}.}
and a typical price for fuel (e.g., gasoline) per liter ($C_{\text{fuel}}$).%
\footnote{We focus on the price of a liter, but this can easily be adjusted for electric vehicles by using electric energy consumption and battery charging cost (cf.\ \cref{sec_discussion_electric_vehicles}).}
The cost of the travel time is calculated as the product of the (estimated) travel time for the trip ($\text{time}_{\text{trip}}$) and a monetary value for time spending ($C_{\text{time}}$).
The latter value can be configured by drivers during trip planning via the navigational system, depending on their personal preference.
This can be compared with choosing a route from multiple options such as \emph{fast} or \emph{economic} that are provided by the navigational systems already today.

The desired level of fuel-efficient driving, as well as the opportunity cost for an extended trip time, depends on the individual drivers' personal preferences~\cite{dai2015personalized}.
Based on the nominal values of $C_{\text{fuel}}$ and $C_{\text{time}}$, the prioritization between fuel consumption and travel time can be adjusted.
This allows for a comparative and real-world applicable evaluation of all kinds of driving systems, including platooning.
%
If required, the formula can be extended with additional monetary terms and utilities that influence a driver's overall trip cost.
These may include platooning-related costs and benefits (e.g., for leading, profit-sharing, maneuver overhead, or toll discounts)~\cite{ledbetter2019lips,johansson2022platoon,earnhardt2022cooperative,malik2023unlocking}, as well as personal utilities (e.g., enjoyment of high speed, satisfaction from reducing fossil fuel consumption, or reduced stress and increased comfort through automation)~\cite{hoffmann2022seizing,sturm2020taxonomy}.
Subjective aspects like comfort can be approximated through measurable proxies such as driving smoothness or time spent as a follower within a platoon.
These proxies could be integrated as soft constraints (e.g., a minimum required resting period) or incorporated as monetized utility terms, using appropriate personal valuation factors.

\subsection{Time Monetization based on Real-World Statistical Data}%
\label{sec_metric_time_mapping_real}


\noindent 
A key component of our cost metric is the driver's personal valuation of time, representing the opportunity cost associated with time spent traveling.
While trip duration does not incur direct monetary costs like fuel consumption, it entails an implicit opportunity cost that has to be ``paid'' through the driver's time investment.
While drivers may subjectively choose values for their personal valuation of time on a per-trip basis, we require a consistent method for assigning these values throughout this paper to enable a systematic analysis of the total trip cost metric.

The opportunity cost of time is difficult to quantify directly, as it varies across individuals and even for the same individual depending on their outside options.
In principle, opportunity costs are an established economic category trying to capture the resources in general (e.g., costs, time) that one spends for choosing one option so that another option must be missed~\cite{parkin2016opportunity}.
They are, in that sense, the lost profit for not using the second option.
This is useful as the first option's competitive advantage over the second option depends not only on the first option's utility but also on the difference to the second option.

With respect to time preferences, opportunity costs are a useful measurement entity.
In case a driver's opportunity cost is very low, the driver would probably spare more time for completing the trip.
A corresponding lower driving speed would require less fuel and thus less actual cost.
If, in contrast, a driver's opportunity cost is very high, the driver would probably try to finish the trip as fast as possible in order to become productive again after leaving the vehicle.
The corresponding higher driving speed would, of course, require more fuel and thus more actual cost.

There is a good rationale to assume that time spent for traveling is directly linked to economic productivity.
The time to travel a fixed distance is correlated with productivity on a global level:
The higher the productivity of a country and the according wages, the faster people tend to travel a given distance~\cite{levine1999pace,bettencourt2007growth}.
The same correlation also holds true on the individual level:
High individual productivity is associated with the strong tendency to spend less time traveling and opt for time saving~\cite{goldbach2020fast,hoffmann2021hare}.
Similarly, highly productive humans (measured in income) are less likely to have patience for spending their time waiting in line~\cite{holt2023examining}.
%
Individuals' value of time is likely very heterogeneous as various groups value their time differently (see also~\cite{farokhi2013game-theoretic}).
While some individuals could earn a lot of money if they could be productive at work, others could not.
To phrase it differently, the drivers could ask themselves \emph{``what would I earn during the time I (still) need for completing my trip if I had reached my destination already?''}
This time value disparities between individual drivers are best measured in their opportunity costs of wages, reflecting their productivity (cf.\ \cite{dong2024optimization}).

\begin{figure}
\centering
\includegraphics[width=\columnwidth]{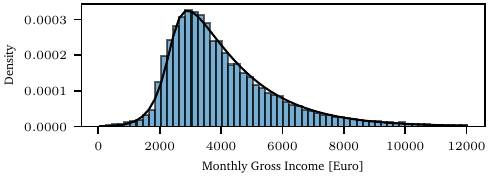}%
\caption{%
Distribution (and fit) of monthly gross income (in Euro) of German full-time workers in April 2023.\protect\footref{ft_brutto_income_source}
}%
\label{fig_brutto_income_fit}
\end{figure}

We therefore consider a simple yet traditional and intuitive mapping from time to a monetary value by looking at peoples' salary as proposed in our earlier work~\cite{heinovski2024platooning}.
The finding that income and value of time are correlated is well established~\cite{fosgerau2005speed,rienstra1996speed,vanommeren2006optimal,hoffmann2021hare}.
Specifically, we use real-world statistical data for the typical monthly gross income of German full-time workers in April 2023.\footnote{%
\label{ft_brutto_income_source}%
Statistisches Bundesamt (Destatis) 2024, statistics available at
\url{https://www.destatis.de/DE/Themen/Arbeit/Verdienste/_Grafik/_Interaktiv/verteilung-bruttomonatsverdienste-vollzeitbeschaeftigung.html}, last accessed: October 23, 2024.
}
We fit a generic hyperbolic distribution from scipy\footnote{\url{https://docs.scipy.org/doc/scipy/reference/generated/scipy.stats.genhyperbolic.html}, version \texttt{1.10.1}} shown in \cref{fig_brutto_income_fit} to the data using the following parameters:
\begin{equation}\label{eq_fitting_parameters}%
\begin{split}
a \approx 0.62 \text{~,~} b \approx 0.39 \text{~,~} p \approx 1.23\\
l \approx 2498.26 \text{~,~} s \approx 363.96 \quad \text{,}
\end{split}
\end{equation}
where
$a$ determines the shape,
$b$ the skewness,
$p$ the tail,
$l$ the location,
and $s$ the scale.


To use the data, we sample values from the fitted distribution to assign a monetary value to drivers' time (cost) proportional to their income.
As shown in \cref{fig_brutto_income_fit}, the values are in \SIrange{0}{12000}{\eu} with a mode of \SI{3000}{\eu}, median of \SI{3738}{\eu}, and mean of \SI{4147}{\eu}.
Since the data and thus the samples from our fitted distribution are values for the monthly income, we divide a sample by \num{160} to transform it into an hourly value, which we use as the cost per hour travel time.
This results in a distribution of \SIrange{0}{75}{\eu} with a mode of \SI{18.1}{\eu}, median of \SI{23.1}{\eu}, and mean of \SI{25.7}{\eu}.
Note that the original fitted distribution produces values larger than \SI{12000}{\eu}, which is the maximum value in the original data source.
Therefore, we limit the samples to this value and the corresponding hourly samples to \SI{75}{\eu}, resulting in a small and thus negligible bias towards the right edge of the distribution.
We assume that drivers would choose a desired driving speed in correlation to their respective monetary value for time, such that their personal preference is reflected in the expected travel time.


\subsection{Understanding the Metric: Numerical Analysis}%
\label{sec_metric_numerical}

\noindent 
To illustrate the concept of our metric, we perform numerical simulations for an abstract scenario:
We consider a single vehicle driving a fixed \SI{50}{\km} trip on an abstract empty freeway.
The driver can choose from various monetary values for $C_{\text{time}} \in \left[ 0,75 \right]\si{\eu}$ as cost per hour.
Empirical data shows a consistent and significant correlation between income and desired speed~\cite{fosgerau2005speed,rienstra1996speed,vanommeren2006optimal,hoffmann2021hare}.
These studies suggest that individuals with higher incomes tend to travel faster (e.g., elasticity of about \num{0.2}~\cite{fosgerau2005speed}), thereby reflecting a higher valuation of time.
To capture this effect, we model the relationship between the time cost $C_{\text{time}}$ and desired driving speed of the vehicle $v_{\text{desired}}$ using a linear function as follows:
\begin{equation}\label{eq_metric_numerical_speed_mapping}%
v_{\text{desired}} = \frac{\SI{55}{\mps} - \SI{22}{\mps}}{\SI{75}{\eu} - \SI{0}{\eu}} \cdot C_{\text{time}} + \SI{22}{\m\per\s} \quad \text{.}
\end{equation}
Thus, it is in $\left[ 22,55 \right]\si{\mps}$ (roughly $\left[ 80, 200 \right]\si{\kmh}$) with an average of \SI{130}{\kmh} and mode of \SI{120}{\kmh}.\footnote{%
The parameters roughly correspond to typical traffic observed on a German Autobahn with a recommended driving speed of \SIrange{120}{130}{\kmh} and a wide range of driving behaviors -- from slow moving vehicles and trucks (\SI{80}{\kmh}) to high-speed passenger vehicles (\SI{200}{\kmh}).%
}
%
We model the fuel consumption by using the \ac{HBEFA}\footnote{\url{https://www.hbefa.net/}} version 3.1, following the approach implemented in SUMO and used in our earlier work~\cite{heinovski2024where}.
In particular, we use values from the \textit{PC\_G\_EU4} emission class, which represents a gasoline-driven passenger car with an engine corresponding to the European norm version 4.
Thus, we can calculate the vehicle's instantaneous fuel consumption rate given its driving speed and acceleration.

In addition to human driving, we simulate platooning by assuming the same vehicle is part of a hypothetical platoon.
Applying the approach from our earlier work~\cite{heinovski2018platoon,heinovski2024where}, the vehicle experiences a constant reduction in fuel consumption of roughly \SI{12}{\percent}%
\footnote{%
According to \textcite{bruneau2017flow}, the position of the vehicle within a platoon, its length, and the gap to the preceding vehicle determine the air drag reduction.
Here, we assume that the vehicle is driving in the middle of the (hypothetical) platoon, which leads to a corresponding air drag reduction of \SI{27}{\percent} (cf.\ \cite[Table 5]{bruneau2017flow}).
\textcite{sovran1983tractive} defines a correlation between the change of the air drag and the change of the fuel consumption (\SI{46}{\percent}), which we utilize to calculate the overall reduction in fuel consumption of \SI{12.42}{\percent}.%
}
due to the slipstream effect as a follower.
%
We model the compromise in driving speed that vehicles often have to make when driving in a platoon by adjusting the vehicle's desired speed by a constant value in $\left[ -5, 5 \right]\si{\mps}$, albeit limited to the maximum driving speed of \SI{55}{\mps}.
Thereby, we are able to show the impact of driving at slightly different speeds than desired.
In the following, we will be calling this updated driving speed the \emph{speed adjustment}.

\subsubsection{Time-Cost-Speed Dependencies}%
\label{sec_metric_numerical_time_cost_speed}

\begin{figure}
\centering
\includegraphics[width=\columnwidth]{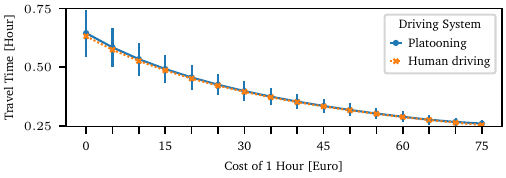}%
\caption{%
    Average travel time for human driving and platooning.
}%
\label{fig_analytical_travel_time}
\end{figure}

\Cref{fig_analytical_travel_time} shows the average total travel time for human driving and platooning with respect to various values for $C_{\text{time}}$ as cost per hour.
%
Since the travel time is inversely proportional to the driving speed, it decreases with an increasing value of $C_{\text{time}}$ for both approaches.

\begin{figure}
\centering
\includegraphics[width=\columnwidth]{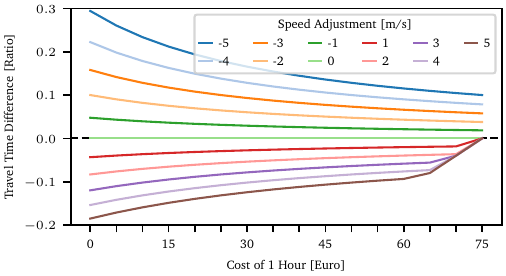}%
\caption{%
    Difference in travel time of platooning to human driving for various speed adjustments.
}%
\label{fig_analytical_travel_time_diff}
\end{figure}

The impact of the speed adjustments in platooning is shown in \cref{fig_analytical_travel_time_diff}.
A negative time difference represents time savings compared to driving without platooning.
The impact is relatively large at low time costs (low absolute driving speeds).
Here, an absolute difference of \SI{5}{\mps} changes the speed already by \SI{23}{\percent} at \SI{22}{\mps} ($C_{\text{time}} = \SI{0}{\eu}$).
Additionally, negative adjustments have a larger impact than positive ones, leading to a slightly increased average travel time.
For high time costs, the impact is less dominant and savings are less likely due to a bounded maximum speed.


\begin{figure}
\centering
\includegraphics[width=\columnwidth]{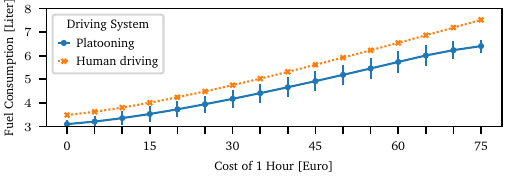}%
\caption{%
    Average fuel consumption for human driving and platooning.
}%
\label{fig_analytical_fuel_consumption}
\end{figure}

\Cref{fig_analytical_fuel_consumption} shows the fuel consumption for human driving and platooning.
%
The average fuel consumption increases with an increasing value of $C_{\text{time}}$ for both approaches.
In platooning, the fuel consumption is generally lower due to the slipstream effect.
Due to the constant \SI{12}{\percent} reduction, the absolute impact is more dominant at higher absolute driving speeds.

\begin{figure}
\centering
\includegraphics[width=\columnwidth]{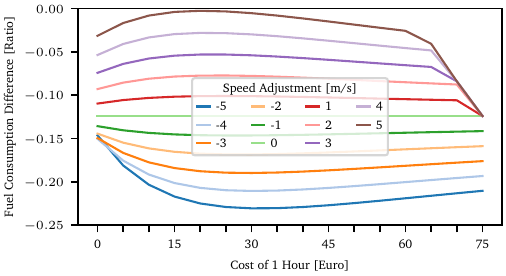}%
\caption{%
    Difference in fuel consumption of platooning to human driving for various speed adjustments.
}%
\label{fig_analytical_fuel_consumption_diff}
\end{figure}

To give more insights into the fuel savings in platooning, we report the difference in fuel consumption of platooning to human driving for various speed adjustments in \cref{fig_analytical_fuel_consumption_diff}.
The relative difference in fuel consumption is negative for all adjustments, indicating that platooning uses less fuel in all cases regardless of the speed deviation.
Naturally, no adjustment leads to the constant saving of \SI{12}{\percent}.
In general, negative adjustments (i.e., driving slower than desired) require less fuel and thus lead to larger savings and vice versa for positive adjustments.

However, some interesting effects following the value of $C_{\text{time}}$ can be observed.
At low values for $C_{\text{time}}$ (i.e., \SIrange{0}{20}{\eu}), an increase of the time cost combined with a positive speed adjustment leads to a reduction in fuel savings due to the faster driving speed, while a negative adjustment leads to an increase due to slower driving.
%
At $C_{\text{time}} = \SI{20}{\eu}$, we can observe the minimum fuel savings for faster driving due to positive speed adjustments (e.g., \SI{5}{\mps} is almost equal to human driving).
On the other hand, for slower driving due to negative speed adjustments, we observe the maximum fuel savings at $C_{\text{time}} = \SI{30}{\eu}$.
%
At medium to high values for $C_{\text{time}}$ (i.e., \SIrange{30}{75}{\eu}), the driving speed increases even further following the time cost value.
Now, the non-linear growth of the fuel consumption with driving speed leads to a significantly higher absolute consumption at higher speeds for both approaches.
For positive speed adjustments, the absolute fuel consumption is larger, but the constant \SI{12}{\percent} reduction due to the slipstream effect also leads to larger absolute savings.
%
In contrast, when looking at the negative speed adjustments, we can observe that driving slower than desired becomes less and less beneficial with increasing desired speed (time cost) as fuel savings decrease.


\subsubsection{Total Trip Cost}%
\label{sec_metric_numerical_trip_cost}

\begin{figure}
\centering
\includegraphics[width=\columnwidth]{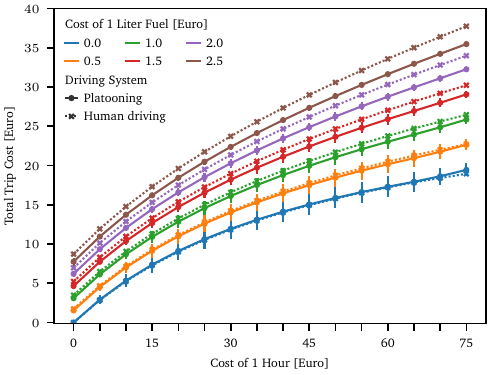}%
\caption{%
    Total trip cost $C_{\text{trip}}$ (our proposed metric, see \cref{eq_total_trip_cost_metric}) for human driving and platooning (average with standard deviation for various speed adjustments).
    The markers distinguish human driving and platooning.
    The color indicates various values for $C_{\text{fuel}}$ (cost per liter).
}%
\label{fig_analytical_total_trip_cost}
\end{figure}

We now investigate the average total trip cost $C_{\text{trip}}$ (our proposed metric) as a function of $C_{\text{time}}$.
We show results for various values of $C_{\text{fuel}}$ as cost per liter fuel in \cref{fig_analytical_total_trip_cost}.
Naturally, the total trip cost increases with the time cost $C_{\text{time}}$ as well as the fuel cost $C_{\text{fuel}}$.
Platooning generally has lower (or equal) trip costs on average, and the difference to human driving becomes larger at higher time and fuel costs.

At $C_{\text{time}} = \SI{0}{\eu}$, the travel time is neglected in the calculation of the total trip cost such that it only depends on the specific fuel cost $C_{\text{fuel}}$.
The average total trip cost of platooning is slightly higher than with human driving due to more negative speed adjustments and the limited driving speed at the largest time cost values.
The relationship of the fuel cost $C_{\text{fuel}}$ to $C_{\text{trip}}$ is straightforward, but the impact of increasing the time cost $C_{\text{time}}$ is not immediately intuitive:
While the travel time decreases with larger time cost values (see \cref{fig_analytical_travel_time}), the total trip cost increases, but with a decreasing slope.
This is due to the travel time becoming a more significant part of the total trip cost metric with larger values of $C_{\text{time}}$.
Thus, driving even faster at already high speeds has less benefit than driving faster at slow speeds but can be beneficial at low to medium time costs.
For better understanding, let us consider the curve for $C_{\text{fuel}} = \SI{0}{\eu}$, where only travel time is considered in the total trip cost.
The time cost $C_{\text{time}}$ is included in both the total trip cost metric as well as the travel time due to how the driving speed is chosen (see \cref{eq_metric_numerical_speed_mapping}).
Thus, the trip cost metric becomes a rational function (a ratio of a linear term and a sum), which leads to a non-linear relationship between the time cost and the total trip cost, which grows faster than the time cost itself.
A constant curve, for instance, would indicate that the total trip cost and the driving speed (given by the time cost) are perfectly balanced, leading to no change in the combined result, e.g., doubling the speed would half the trip cost (determined by the travel time), which is not the case.

If the fuel cost is considered ($C_{\text{fuel}} > \SI{0}{\eu}$) in the trip cost metric, larger values decrease the impact of the travel time and thus the aforementioned effect, making the curve more steep even for higher time cost values.
Thus, saving fuel by driving slower becomes more relevant, especially at large fuel costs.
%
However, especially at higher values, the time cost ($C_{\text{time}}$) remains the dominant factor across the considered parameter range ($C_{\text{time}} \in \left[0, 75\right]\si{\eu}$, $C_{\text{fuel}} \in \left[0, 2.5\right]\si{\eu}$).
This highlights the implicit weighting between time cost and fuel cost in our metric.

\section{Incentive-based Platoon Formation}%
\label{sec_platoon_formation}

\noindent 

\subsection{Assumptions}%
\label{sec_platoon_formation_assumptions}

\noindent 
We focus on spontaneous individual traffic of passenger cars, where drivers perform unsynchronized uncorrelated trips on freeways. 
We assume drivers can freely customize their trip based on an individually preferred traveling speed during trip planning.
Vehicles will start their trips driving individually but immediately start searching for appropriate platooning opportunities en route.
We do not assume that platooning is generally preferred and instead consider driving individually (i.e., with \ac{ACC}) and platooning as equally valid options.
The trips and drivers' requirements are not known beforehand and, thus, no pre-planning of platoon configurations can be realized.
We assume that vehicles can collect information about platoons and other vehicles by means of 4G/5G-based \ac{C-V2X} or \p{}-based \ac{DSRC}.
Similarly, vehicles can exchange information for maneuver control and platoon operation.
Therefore, the platoon formation, including (1) computation of \vtp{} assignments and (2) performing corresponding join maneuvers, happens on-demand and en route.

\subsection{Platoon Formation Algorithm}%
\label{sec_platoon_formation_algorithm}

\begin{algorithm}
    \caption{Heuristic for Trip Cost-based Platoon Formation}%
    \label{alg_formation}
    \begin{algorithmic}
        \REQUIRE{list of (available) platooning opportunities in range}
        \STATE{estimate cost for the remaining trip driving individually;}
        \FORALL{platooning opportunities}
            \STATE{define case for join maneuver;}
            \STATE{estimate distance \& cost for join maneuver;}
            \STATE{calculate distance shared with platoon;}
            \STATE{estimate cost for distance shared with platoon;}
            \IF{distance still remaining > 0}
                \STATE{estimate cost for remaining distance;}
            \ENDIF{}
            \IF{sum of costs for platooning < cost for individual driving}
                \STATE{consider opportunity as feasible;}
            \ENDIF{}
        \ENDFOR{}
        \IF{feasible opportunities}
            \STATE{select opportunity with minimum cost;}
            \STATE{trigger corresponding join maneuver;}
        \ENDIF{}
    \end{algorithmic}
\end{algorithm}

\begin{figure}
\centering
    \begin{tikzpicture}
    \tikzstyle{every node}=[font=\small]
    \node[draw,minimum width=\columnwidth,text centered,text depth=5em] (estimation_p) {Cost Platoon Driving (\cref{sec_platoon_formation_estimation_platoon})};
    \node[draw,text width=6em,text centered] (estimation_p_join) at ([yshift=-.75em,xshift=4.5em]estimation_p.west) {Cost Join Maneuver (\cref{sec_platoon_formation_estimation_join})};
    \node[draw,text width=6em,text centered] (estimation_p_shared) at ([xshift=4.75em]estimation_p_join.east) {Cost Shared Distance (\cref{sec_platoon_formation_estimation_shared})};
    \node[draw,text width=6em,text centered] (estimation_p_remaining) at ([xshift=4.75em]estimation_p_shared.east) {Cost Remaining Distance (\cref{sec_platoon_formation_estimation_remaining_individual})};
    %
    %
    \node[draw,minimum width=\columnwidth,text centered,text depth=.5em,] (estimation_i) at ([yshift=1.5em]estimation_p.north) {Cost Individual Driving (\cref{sec_platoon_formation_estimation_individual})};
\end{tikzpicture}
\caption{%
Overview of all cost estimation parts of our algorithm.
}%
\label{fig_algorithm_estimation_overview}
\end{figure}
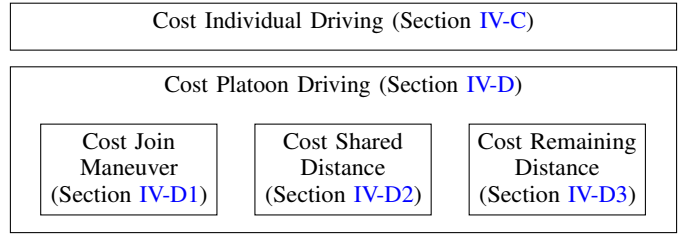

\noindent 
A greedy formation algorithm is periodically executed on the vehicle in order to evaluate available platooning opportunities.
Our recent work showed that such greedy heuristics perform very well in comparison to a global optimum, and are much faster in terms of finding a suitable platoon~\cite{heinovski2024where}.
Forming platoons always consists of the following steps:
(1) data collection of available vehicles and platoons,
(2) computation of vehicle-to-platoon assignments depending on the selected approach,
(3) execution of join maneuvers to implement the computed assignments.
%

Using information about nearby vehicles, every vehicle performs the heuristic given in \cref{alg_formation} for forming an appropriate platoon.
For every available platooning opportunity, the algorithm estimates the total cost for the remaining trip under the assumption of joining the target platoon (or forming a new platoon), and compares it to the (estimated) cost of driving individually (i.e., with \ac{ACC}), using our total trip cost metric from \cref{sec_metric}.
Thereby, we consider (cumulative) quantitative benefits of \ac{ACC} and platooning, instead of (instantaneous) qualitative aspects such as (current) deviation in driving speed (cf.\ \cite{%
hoshino2013reactive, 
su2016autonomous, 
heinovski2024where, 
}).
If a vehicle cannot find a feasible platooning opportunity, it continues to drive alone using \ac{ACC}.
%
Feasibility is determined only by comparing the trip costs and independent of a platooning opportunity's deviation in (desired) driving speed (cf.\ \cite{heinovski2024where}).
The trade-off between travel time and fuel consumption given by the driving speed is done exactly by applying the trip cost metric.
An upper limit for the deviation would artificially constrain platooning options, making the algorithm less flexible.
An overview of all parts of the estimation including pointers to corresponding sub-sections is shown in \cref{fig_algorithm_estimation_overview}.

Among all feasible platooning opportunities with a lower estimated cost than driving alone, our algorithm picks the one with the smallest cost and triggers a corresponding join maneuver.
Vehicles can join other vehicles or already existing platoons at the front or at the back, thereby becoming the new platoon leader or the new last member, respectively.
A joining vehicle has to close the gap to the target platoon by either speeding up or slowing down, depending on its position, and adjust its speed to the speed of the platoon.
Vehicles consider the overall maneuver time under ideal conditions, i.e., neglecting potential interference due to traffic conditions, for simplicity.
Solutions for coping with other vehicles interfering the maneuvers have been proposed~\cite{segata2014supporting,strunz2021coop,paranjothi2020pmcd}, but are not in the focus of this work.

After successful completion of the join maneuver, vehicles are driving with \ac{CACC} and stay in the platoon as long as possible, i.e., until they reach their destination or all other platoon members leave.%
\footnote{%
In this work, we focus on a single platoon assignment in order to understand the effects of the speed trade-off in comparison to our recent work~\cite{heinovski2024where}.
However, our approach using the total trip cost metric can also be applied to continuous evaluation of available platooning opportunities that potentially leads to switching to a better platoon.
}
Assuming a fully operational \aca{CACC}, vehicles in a platoon always mirror the behavior of the platoon leader and keep a constant gap of \SI{5}{\m}~\cite{rajamani2000demonstration,jootel2012final}.
If the platoon leader leaves the platoon, the next remaining vehicle within the formation becomes the new leader, keeping all properties of the platoon.
If all other platoon members leave the platoon, the vehicle continues to drive individually and starts searching for new platooning opportunities.

\subsection{Estimating the Trip Cost for Individual Driving}%
\label{sec_platoon_formation_estimation_individual}

\noindent 
On every execution of the formation algorithm, it first estimates the cost for the remaining trip distance assuming individual driving with \ac{ACC}.
This cost serves as a baseline for selecting the platooning opportunities resulting in cost saving. 
%
Using the vehicle's desired driving speed defined by the driver's time cost value $C_{\text{time}}$, we first estimate the remaining travel time.
Then, the fuel model from our earlier work~\cite{heinovski2024where} (cf.\ \cref{sec_metric_numerical}) is used to estimate the fuel consumption rate for driving at the desired speed.
%
Using the estimated fuel consumption (in liters) and the estimated travel time (in hours), we calculate the estimated trip cost using our total trip cost formula given in \cref{eq_total_trip_cost_metric}.

\subsection{Estimating the Trip Cost for Platoon Driving}%
\label{sec_platoon_formation_estimation_platoon}

\noindent 
The total cost for platooning consists of the following parts:
(1) the cost for a corresponding join maneuver (see \cref{sec_platoon_formation_estimation_join}),
(2) the cost for the part of the trip that can be driven in the target platoon,
and (3) the cost for the remaining part of the trip after all other platoon members left (see \cref{sec_platoon_formation_estimation_remaining_individual}).
Of course, these costs are a function of the quality of the \vtp{} assignment algorithm.
%
If the total of these costs is lower than the cost of individual driving, the corresponding platooning opportunity is considered feasible.
Thereby, we make sure that driving individually (i.e., with \ac{ACC}) always stays a valid option if beneficial in terms of the total cost for the remaining trip, not enforcing platooning like pure similarity-based approaches (cf.\ \cref{sec_related_work_traditional}).
Additionally, we are able to balance the trade-off between a well-fitting platooning opportunity and a lengthy join maneuver, which pure similarity-based approaches can not.

\subsubsection{Join Maneuver}%
\label{sec_platoon_formation_estimation_join}

\noindent 
While estimating the trip cost including travel time and fuel consumption for a fixed distance and a given desired speed is straightforward, estimating the cost for a join maneuver is more difficult.
%
%
We model every join maneuver case by the following steps:
(1) adjusting the current speed of the joiner to a constant speed for approaching the target platoon (closing the gap),
(2) closing the gap to the target platoon by driving at this constant speed,
and (3) adjusting the current speed of the joiner to the current speed of the target platoon.
Depending on the specific situation, the joiner will need to accelerate or decelerate in steps (1) and (3).
%
We generally assume that maneuvers will not be performed as fast as possible by always accelerating to the maximum possible speed for approaching, even if the joining vehicle is already faster than the target platoon.
Instead, the constant speed for closing the gap is defined based on the current driving speed of the target platoon and a fixed coefficient (e.g., \SI{15}{\percent}).
However, we enforce at least a \SI{1}{\mps} speed difference between the joiner and the target platoon to finish a join maneuver in reasonable time.

First, we estimate the distance and time required for steps (1) and (3), and use the result together with the estimated distance the target platoon will be driving during the join maneuver to estimate step (2).
If the gap between joining vehicle and target platoon can be closed already in step (1) or in steps (1) \& (3) combined, step (2) is skipped.
Second, we estimate the fuel required for each step as the product of time and fuel rate at either constant acceleration \& deceleration (with average speed of the steps (1) and (3)), or constant speed (with \num{0} acceleration).
Finally, we sum up the distance, time, and fuel for all steps and estimate the total cost for the join maneuver by applying our trip cost metric.

\subsubsection{Shared Distance Driving in Platoon}%
\label{sec_platoon_formation_estimation_shared}

\noindent 
After estimating the distance driven during the join maneuver, we calculate the part of the remaining trip that the vehicle can actually drive in the target platoon.
We call this the \emph{shared distance}.
If all other (already existing) members of the target platoon reach their destination and, thus, leave the platoon earlier than the searching vehicle reaches its destination, it will need to switch to driving individually.
%
In order to calculate the shared distance, the algorithm first finds the furthest destination of all platoon members.
If this destination is further than the searching vehicle's destination, it can actually share the entire remaining trip with the platoon.
Otherwise, the shared distance is the distance until the furthest destination subtracted by the distance required for the join maneuver.
%
We can now estimate the total cost for the shared distance based on the estimated travel time and fuel consumption.
Instead of the vehicle's desired driving speed, this time, we use the desired driving speed of the platoon.
%
For the calculation of fuel consumption, we need to consider that vehicles can join a platoon as a new leader at the front or as a new last vehicle at the end.
Thus, we calculate the reduction in fuel consumption (and emissions) based on their position within the platoon~\cite{bruneau2017flow} after joining.
We apply the constant reduction (either \SI{5}{\percent} as leader or \SI{11}{\percent} as last vehicle) to the calculated fuel consumption to integrate the slipstream effect in the trip cost estimation.

\subsubsection{Remaining Distance}%
\label{sec_platoon_formation_estimation_remaining_individual}

\noindent 
%
The remaining distance can be calculated by subtracting the shared distance and the distance required for the join maneuver from the distance of the entire trip.
This remaining distance will be driven using \ac{ACC} at the vehicle's original desired driving speed.
Thus, the estimation of the cost for this part of the trip follows the individual driving case (cf.\ \cref{sec_platoon_formation_estimation_individual}).
Note that, even though we assume the vehicle will drive individually for the given part of the trip, this might not be the case later on.
The vehicle might find and join another platoon, profiting again from platooning benefits. 

%

\section{Evaluation}%
\label{sec_evaluation}

\noindent 
We evaluate our proposed trip cost-based platoon formation algorithm by comparing it to an existing, similarity-based \vtp{} assignment algorithm and two baseline approaches without platooning.
In the following, we first present the simulation setup in \cref{sec_simulation_setup}.
We then report results for a time monetization using real-world statistical data for workers' income (see \cref{sec_metric_time_mapping_real}) in \cref{sec_results_real}.
Finally, we present results for an artificial monetization of drivers' time cost value in \cref{sec_results_artificial}.

%

\subsection{Simulation Setup}%
\label{sec_simulation_setup}

\noindent 
We aim to analyze the behavior of the platoon formation algorithms and observe corresponding platooning benefits in a large-scale scenario with many vehicles.
All results presented in the following were obtained in a large-scale simulation study using \simulator{},\footnote{\url{https://www.plafosim.de/}, version \texttt{0.17.3}} which is tailored towards simulation of platoon management rather than microscopic control of the involved vehicles~\cite{heinovski2021scalable}.
In our study, we conduct individual simulations for the following approaches:
\begin{itemize}
\item \human{} -- manual driving (following the Krauss model~\cite{krauss1998microscopic}).
\item \acc{} -- vehicles controlled by \ac{ACC}~\cite[Eq.\ 6.18]{rajamani2012vehicle}.
\item \similaritybased{} -- vehicle platooning (\ac{CACC}) using the \emph{distributed greedy} approach from~\cite{heinovski2024where}.
\item \tripcostbased{} -- vehicle platooning (\ac{CACC}) using our proposed approach (see \cref{sec_platoon_formation}).
\end{itemize}

\begin{table}
    \footnotesize
    \centering
    \caption{Simulation parameters for road and traffic.}%
    \label{tab_params_traffic}
    \begin{tabular}{lr}
        \toprule
        Parameter                                   & Value \\
        \midrule
        Freeway length                              & \SI{100}{\km} \\
        Number of lanes                             & 3 \\
        Ramp interval                               & \SI{10}{\km} \\
        Depart positions                            & random on-ramp \\
        Arrival positions                           & random off-ramp at trip end \\
        \midrule
        Car Following model                         & Krauss, \ac{ACC}, and \ac{CACC} \\
        Krauss desired headway                      & \SI{1}{\s} \\
        \ac{ACC} desired headway                    & \SI{1}{\s} \\
        \ac{CACC} desired gap                       & \SI{5}{\m} \\
        Max.\ speed                                 & \SI{55}{\mps} (roughly \SI{200}{\kmh}) \\
        Max.\ acceleration                          & \SI{2.5}{\m\per\s\squared} \\
        Max.\ deceleration                          & \SI{10}{\m\per\s\squared} \\
        \midrule
        Fixed trip length                           & \SI{50}{\km} \\
        Cost of \SI{1}{\hour} $C_{\text{time}}$     & according to distribution \\
        Desired speed                               & according to cost of \SI{1}{\hour} $C_{\text{time}}$ \\
        Min.\ desired speed                         & \SI{22}{\mps} (roughly \SI{80}{\kmh}) \\
        Max.\ desired speed                         & \SI{55}{\mps} (roughly \SI{200}{\kmh}) \\
        \midrule
        Desired density                             & \SIlist[list-units=single]{5;10;15;20;25}{\density} \\
        Desired no.\ of vehicles                    & \SIlist[list-units=single]{1500;3000;4500;6000;7500}{\vehicle} \\
        Departure rate                              & \SIlist[list-units=single]{3564;7129;10693;14257;17822}{\flow} \\
        \bottomrule                                 \\
    \end{tabular}
\end{table}

We follow the scenario and general methodology described in our recent work~\cite{heinovski2024where} for better comparability:
We consider a \num{3}-lane freeway of \SI{100}{\km} length with periodic on-/off-ramps every \SI{10}{\km}, which allow vehicles to enter and leave the freeway.
Vehicles perform trips of \SI{50}{\km} between a pair of randomly (equally distributed) selected on-/off ramps.
To abstract from the detailed dynamics of merging from an on-ramp onto the freeway, vehicles are initialized directly at their designated position on the freeway, using their desired driving speed and occupying the right-most feasible lane.
Similarly, the process of merging from the freeway onto an off-ramp is simplified by removing (individually driving) vehicles from the simulation once they reach their assigned destination.
We assume a road network without any disturbances to the road infrastructure (e.g., by road construction) or by traffic accidents.

%
Vehicles start their trips driving individually, using either
the popular Krauss model~\cite{krauss1998microscopic} for human driving
or a standard \ac{ACC}~\cite[Eq.\ 6.18]{rajamani2012vehicle}, the de facto standard for all modern vehicles.
They use a constant time-based safety gap, realistic acceleration \& deceleration limits (see \cref{tab_params_traffic}), and safe maneuvering logic in order to avoid collisions.
All vehicles obey traffic rules, including lane-keeping (i.e., ``keep-right'') and desired speed targets.
%
We model vehicle demand in a macroscopic way:
vehicles are created at a constant insertion (departure) rate in order to maintain a fixed desired traffic density (see \cref{tab_params_traffic} and~\cite[Eq.\ 14-16]{heinovski2024where}).
We do so because we are only interested in a relative comparison between the approaches and not in the maximum possible traffic flow for every approach.
In line with our previous work~\cite{heinovski2024where}, we chose \SI{25}{\density} as highest traffic density as, at higher densities, the scenario is too crowded to insert further vehicles in all simulated approaches.
We assign desired vehicle speeds following the drivers' hourly time cost $C_{\text{time}}$ distribution as described in \cref{sec_metric_numerical}.
We assume that drivers who assess more value to their time also tend to drive faster to reduce the travel time (cf.\ \cref{sec_metric_time_mapping_real}).
%
The desired driving speed is in $\left[ 22,55 \right]\si{\mps}$ (roughly $\left[ 80, 200 \right]\si{\kmh}$) with an average of \SI{130}{\kmh} and mode of \SI{120}{\kmh} (cf.\ \cref{sec_metric_numerical}).
All vehicles use a fixed fuel cost $C_{\text{fuel}} = \SI{1.84}{\eu\per\litre}$, reflecting the typical price of Euro-super \num{95} fuel in Germany during April 2023 -- the reference period for the income data shown in \cref{fig_brutto_income_fit}.\footnote{%
Price with taxes on April 3, 2023 according to \url{https://energy.ec.europa.eu/document/download/906e60ca-8b6a-44e7-8589-652854d2fd3f_en?filename=Weekly_Oil_Bulletin_Prices_History_maticni_4web.xlsx}, last accessed October 23, 2024%
}
This allows a focused evaluation of the impact of personal time valuation, while anchoring fuel cost in a realistic and consistent baseline.
The relationship between time cost and fuel cost is explored analytically in \cref{sec_metric_numerical_trip_cost}.
The fuel consumption of a vehicle is calculated and estimated using the approach described in \cref{sec_metric_numerical}.
A summary of all simulation parameters for the road network and traffic can be found in \cref{tab_params_traffic}.

\begin{table}
    \footnotesize
    \centering
    \caption{Simulation parameters for formation logic.}%
    \label{tab_params_formation}
    \begin{tabular}{lr}
        \toprule
        Parameter                                   & Value \\
        \midrule
        Penetration rate                            & 100\% \\
        Execution interval                          & \SI{60}{\s} \\
        Communication range for advertisements      & \SI{500}{\m} \\
        \midrule
        For \similaritybased{} (cf.\ \cite{heinovski2024where}): \\
        \quad Speed Window $m$ & \num{0.2} \\
        \quad Search range $r$ & \SI{1000}{\m} \\
        \quad Weight of speed vs.\ position $\alpha$ & \num{0.5} \\
        \bottomrule
    \end{tabular}
\end{table}

In both platooning approaches, platoon formation is performed in regular intervals every \SI{60}{\s}. 
We assume that vehicles have information (i.e., desired speed, speed, position, destination, platoon state, and maneuver state) about all other cars within their communication range (\SI{500}{\m}).
Vehicles can join other vehicles or already existing platoons at the front or at the back, thereby becoming the new platoon leader or the new last member, respectively.
After successfully completing the join maneuver, platoon members use a standard \ac{CACC} with constant spacing~\cite[Eq.\ 6]{rajamani2000demonstration} and stay within the platoon until they reach their destination ramp.
Platoon leaders use \ac{ACC}, while followers use a simplified \ac{CACC} implementation, abstracting low-level control, wireless communication details, and platooning maneuvers~\cite{heinovski2021scalable}.
We generally assume robust and string-stable \ac{CACC} operation, as safety aspects such as handling sudden disturbances or system failures are beyond the scope of this work.
Moreover, we assume a penetration rate of \SI{100}{\percent} for platooning capabilities on top of \ac{ACC} in both platooning approaches.%
\footnote{%
While lower penetration rates leading to a mixed-traffic environment are more realistic for the future deployment of platooning, we focus on its effects in large-scale environments.
High penetration rates of platooning have been shown to increase traffic flow, mitigate traffic oscillations, and enhance safety in mixed-traffic scenarios~\cite{peng2025enhancing}.
By assuming a \SI{100}{\percent} penetration rate, our study provides an upper bound on the potential benefits of platooning in future deployments.
Nonetheless, independent of the actual penetration rate, all vehicles can already benefit from \ac{ACC}, and some will additionally benefit from platooning.%
}
A summary of all simulation parameters for the platoon formation can be found in \cref{tab_params_formation}.


We simulate \SI{7200}{\s} (\SI{2}{\hour}) of traffic in multiple repetitions for every approach after pre-filling the road with the desired number of vehicles, using the different (fixed) traffic density values.
We only consider results from vehicles that departed after an initial transient period of \SI{1800}{\s} (\SI{0.5}{\hour}).
We simulate \num{5} runs per density and approach, resulting in a total of \num{100} individual simulation runs for each time monetization configuration.

%

\subsection{Time Monetization using Real-World Data}%
\label{sec_results_real}

\noindent 
We study the performance of our platoon formation algorithm using the trip cost metric to optimize drivers' trip costs.
We first use real-world statistical data for drivers' personal time cost value to achieve realistic results.

\subsubsection{Driving Speed}%
\label{sec_results_real_driving_speed}

\begin{figure*}
\centering
\includegraphics[width=\textwidth]{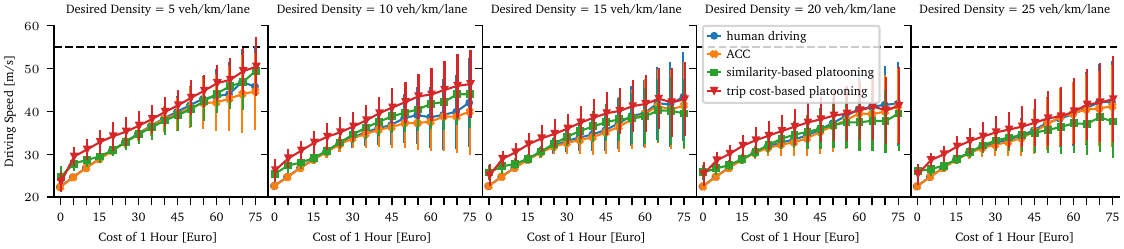}
\caption{%
Average driving speed (with standard deviation) under different traffic densities.
The x-axis shows various values for $C_{\text{time}}$ (cost per hour) based on the income-based distribution (see \cref{sec_metric_time_mapping_real}).
The dashed lines indicate the maximum desired driving speed of \SI{55}{\metre\per\second} (also the maximum possible speed).
}%
\label{fig_results_real_driving_speed}
\end{figure*}


\Cref{fig_results_real_driving_speed} shows the average driving speed (with standard deviation) for
\human{}~(blue),
\acc{}~(orange),
\similaritybased{}~(green),
and \tripcostbased{}~(red)
for different $C_{\text{time}}$ (cost per hour) values and different vehicle densities.
%
Following the configuration of the desired speed and the effects explained in \cref{sec_metric_numerical}, the driving speed increases proportional to the drivers' time cost due to the correlated desired driving speed.
Note that the actual driving speed can differ from the desired speed, since vehicles are influenced by speed adjustments due to platooning decisions as well as by traffic.
In general, we can observe that the speeds of \human{}, \acc{}, and \similaritybased{} are mostly similar, but \tripcostbased{} almost always leads to significantly increased driving speed.
While \human{} and \acc{} can only drive as fast as their desired speed, both platooning approaches actually can exceed their desired speed in order to find (and join) suitable platooning opportunities.
In \similaritybased{}, positive and negative speed adjustments are equally possible as long as the relative deviation is within the defined speed window (i.e., \SI{20}{\percent}).
However, in reality, the deviation in this approach is often negative and larger due to traffic effects.

As can be seen, \similaritybased{} has a lower positive deviation than \tripcostbased{}, resulting also in a lower actual driving speed.
In general, this approach has no upper limit for speed adjustments and thus always tries to maximize the driving speed depending on the time cost value.
If a driver is comfortable only until a certain driving speed, they can use a low time cost value that will favor slower driving in most cases.
An upper limit, similar to the speed window used in \similaritybased{} may improve the driving experience if slower driving is preferred.
However, such a limit would generally constrain platooning options, making the algorithm less flexible and thereby reducing the feasibility of platooning.
On the other hand, a larger speed window for \similaritybased{} increases platooning opportunities but leads to more deviation from the desired driving speed~\cite{heinovski2024where} and can thus lead to increased cost.
The investigation of such speed limits was not the main focus of this work.

When increasing the desired vehicle density in the scenario, the effects of traffic become prominent, and reduce the vehicles' driving speed.
At mid to high densities, \human{} and \acc{} allow for faster driving than \similaritybased{} at large time costs.
The reason is that only a few platooning opportunities exist at these high time cost values, requiring the algorithm to pick slower-driving vehicles and platoons.
In comparison, \tripcostbased{} achieves a higher driving speed under the same condition since this approach allows driving individually if there is no beneficial platooning opportunity.
In general, platooning can cope better with crowded traffic due to many more vehicles being synchronized, which is in line with previously reported results~\cite{heinovski2024where}.

%

\subsubsection{Travel Time}%
\label{sec_results_real_travel_time}

\begin{figure*}
\centering
\includegraphics[width=\textwidth]{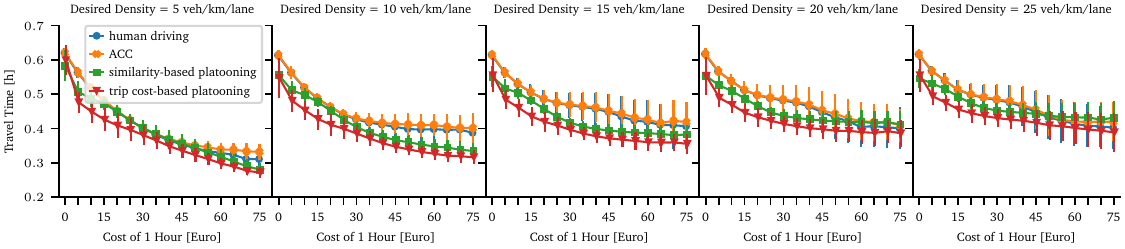}
\caption{%
Average travel time (with standard deviation) under different traffic densities.
The x-axis shows various values for $C_{\text{time}}$ (cost per hour) based on the income-based distribution (see \cref{sec_metric_time_mapping_real}).
}%
\label{fig_results_real_travel_time}
\end{figure*}

\begin{figure*}
\centering
\includegraphics[width=\textwidth]{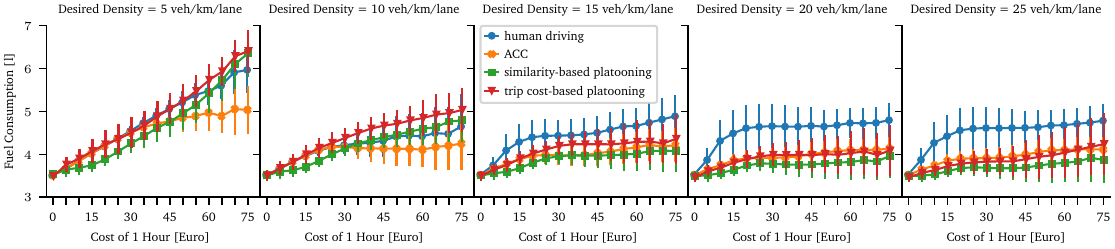}
\caption{%
Average fuel consumption (with standard deviation) under different traffic densities.
The x-axis shows various values for $C_{\text{time}}$ (cost per hour) based on the income-based distribution (see \cref{sec_metric_time_mapping_real}).
}%
\label{fig_results_real_fuel_consumption}
\end{figure*}

\Cref{fig_results_real_travel_time} shows the average travel time (with standard deviation) for
\human{}~(blue),
\acc{}~(orange),
\similaritybased{}~(green),
and \tripcostbased{}~(red)
for vehicles' fixed \SI{50}{\km} trip and different vehicle densities.
The x-axis shows various values for $C_{\text{time}}$ (cost per hour).
The travel time follows the patterns observed from the driving speed
as drivers with larger time costs have shorter travel times due to the higher (desired) driving speed.
Among all time costs and densities, \tripcostbased{} leads to the shortest travel time due to the higher driving speed and the smoother traffic flow.
Using \ac{ACC} leads to the largest travel time due to a decreased driving speed of all vehicles being synchronized in dense traffic.
The slope of the travel time curve becomes less steep with increasing density.
This indicates that higher demands for driving speed (given by the time cost) become less effective if the traffic increases, which we have observed already in \cref{fig_results_real_driving_speed} and \cref{sec_metric_numerical_time_cost_speed}.
%
While platooning generally leads to the shortest travel time, \similaritybased{} achieves worse results than \human{} and \acc{} for high time costs at the largest densities.

%

\subsubsection{Fuel Consumption}%
\label{sec_results_real_fuel_consumption}

We show the average fuel consumption (with standard deviation) for \human{}~(blue), \acc{}~(orange), \similaritybased{}~(green), and \tripcostbased{}~(red) in \cref{fig_results_real_fuel_consumption} for vehicles' fixed \SI{50}{\km} trip and different vehicle densities.
The x-axis shows various values for $C_{\text{time}}$ (cost per hour).
A larger value for $C_{\text{time}}$ in general increases the fuel consumption, while a higher vehicle density in general reduces the fuel consumption due to reduced speed in traffic.
For human driving, starting at \SI{15}{\density}, the traffic is not in free flow mode anymore as vehicles begin to significantly influence each other's driving behavior (cf.\ \cite{forster2014cooperative}).
Under these conditions, gaps between vehicles approach the desired safety gap, causing frequent acceleration \& deceleration maneuvers, which in turn increase fuel consumption.
Synchronized driving using \ac{ACC} already helps reducing fuel consumption, but also suffers from reduced driving speed due to traffic effects (cf.\ \cref{fig_results_real_driving_speed}).
%
For platooning approaches, fuel consumption (also related to driving speed) is influenced by traffic effects and platoon formation.
Vehicles using \similaritybased{} almost always consume less fuel than with \tripcostbased{}, and often also less than \human{} and \acc{}.
The reason is, \similaritybased{} allows slower driving based on the speed of other vehicles in the target platoon, whereas \tripcostbased{} often leads to a faster driving as long as it improves the overall trip cost.
Note that \tripcostbased{} generally outperforms \acc{} at the highest density although it is driving faster (cf.\ \cref{fig_results_real_driving_speed}).

%

\subsubsection{Total Trip Cost}%
\label{sec_results_real_trip_cost}

\begin{figure*}
\centering
\includegraphics[width=\textwidth]{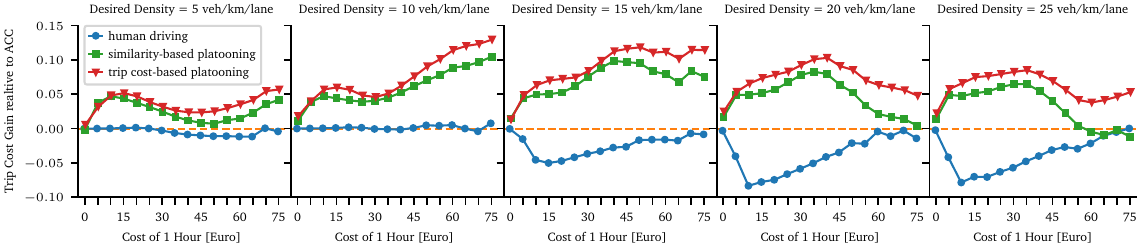}
\caption{%
Average gain on trip cost $C_{\text{trip}}$ relative to \acc{}.
Positive values indicate a trip cost $C_{\text{trip}}$ lower than \acc{}.
The x-axis shows various values for $C_{\text{time}}$ (cost per hour) based on the income-based distribution (see \cref{sec_metric_time_mapping_real}).
}%
\label{fig_results_real_acc_gain}
\end{figure*}

\begin{figure*}
\centering
\includegraphics[width=\textwidth]{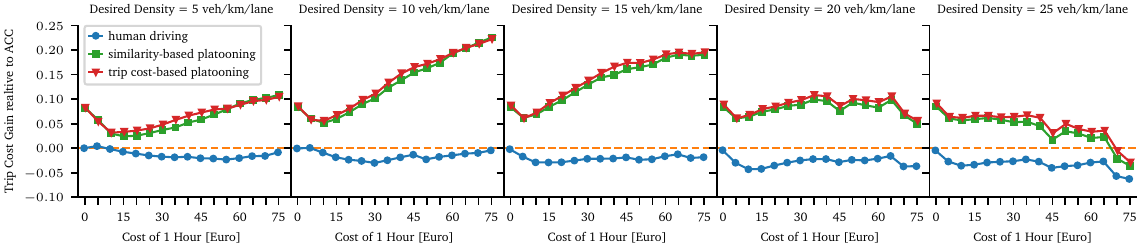}
\caption{%
Average gain on trip cost $C_{\text{trip}}$ relative to \acc{}.
$C_{\text{time}}$ (cost per hour) values are sampled from a bathtub distribution.
}%
\label{fig_results_artificial_acc_gain_bathtub}
\end{figure*}

\Cref{fig_results_real_acc_gain} shows the average gain in total trip cost $C_{\text{trip}}$ (as defined by our metric in \cref{eq_total_trip_cost_metric}) for
\human{}~(blue),
\similaritybased{}~(green),
and \tripcostbased{}~(red),
relative to \acc{}~(orange).
A positive gain indicates a reduction in total trip cost $C_{\text{trip}}$, while a negative value reflects an increase in total trip cost relative to \acc{}.
%
As expected, we can observe that both platooning approaches generally lead to a positive gain in comparison to \acc{}, while \human{} almost always has no or even a strongly negative gain.
Our proposed formation algorithm used in \tripcostbased{} consistently yields the highest overall gain, achieving improvements of up to \SI{13}{\percent} in total and outperforming the \similaritybased{} approach by as much as \SI{6}{\percent}.
This is caused by the higher driving speed in combination with the slipstream effect in platooning.

When fuel consumption is prioritized over time (lowest values of $C_{\text{time}}$), all approaches perform rather similarly.
This is due to the fact that only a few such cars exist at corresponding time cost values according to the income distribution.
As a result, these vehicles drive mostly alone. 
At medium time cost values, we observe a \emph{valley} in the gain curves for both platooning approaches due to an initial decrease followed by an increase.
According to the distribution used for the time monetization, most of the vehicles sample a medium value for $C_{\text{time}}$, i.e., in \SIrange{15}{40}{\eu}.
Hence, the desired speed remains rather low and demands are rather similar such that vehicles can achieve a smooth traffic flow using \ac{ACC}.

When further increasing the time cost, generally \acc{} leads to a lower driving speed than platooning approaches, resulting in longer travel time.
For high densities, this effect is reversed as platooning helps increase road capacity and fewer traffic-related effects are observed.
At the largest density, \tripcostbased{} becomes again increasingly beneficial at high time cost values as several platooning opportunities exist also for high speed demands and the algorithm.
In contrast, the strict platooning requirement of \similaritybased{} leads to joining mostly slower driving vehicles, resulting in a negative gain in comparison to \acc{}.

\subsection{Artificial Time Monetization}%
\label{sec_results_artificial}

\noindent 
To demonstrate the general applicability and benefits of our proposed platoon formation algorithm, we now present results based on a different, more artificial time-cost distribution.
While we experimented with multiple distributions, we found the bathtub distribution particularly interesting.
It represents a scenario where all monetary interest values are generally present, but the extremes are emphasized:
(a) those who place minimal importance on travel time,
and (b) those who prioritize travel time above all else.

\Cref{fig_results_artificial_acc_gain_bathtub} shows the results again in form of the average gain in total trip cost $C_{\text{trip}}$ for
\human{},
\similaritybased{},
and \tripcostbased{},
relative to \acc{}.
Platooning starts with substantial gains at low time cost values due to the abundance of vehicles, which presents ample opportunities for forming platoons.
However, these gains quickly diminish at mid-range time cost values due to the reduced number of vehicles available.
For higher cost values, the benefit increases significantly.
This is due to the platooning gain both in terms of faster driving as well as the slipstream effect.
Here, \tripcostbased{} achieves slightly improved gains compared to \similaritybased{} due to the absence of a speed window, allowing for more feasible platooning opportunities.
Only for high traffic densities, the effect is diminished as the road occupancy is too high for significant gains.
Nonetheless, these results illustrate that our approach functions even in an edge-case scenario with extreme values for drivers' time valuation.

\section{Discussion}%
\label{sec_discussion}

\noindent 
In this section, we discuss relevant remarks and limitations of our proposed \tripcostbased{} approach.

\subsection{Assumptions on \ac{CACC} Stability and Safety}%
\label{sec_discussion_assumptions}

In this work, we assume robust, string-stable \ac{CACC} operation on an idealized, disturbance-free freeway (see \cref{sec_simulation_setup}).
These simplifying assumptions allow us to focus on the core effects of platoon assignment and formation strategies.

Stable and efficient \ac{CACC}-based platooning, such as under constant spacing policies~\cite{rajamani2000demonstration}, is technically feasible if key conditions are met~\cite{feng2019string,locigno2022cooperative}.
String stability requires that vehicles exchange accurate state data (e.g., speed, acceleration) with neighbors -- typically the platoon leader and immediate predecessor -- at sufficiently high frequency~\cite{locigno2022cooperative}.

In real-world settings, communication may suffer from issues such as congestion, delay, or message loss.
To enhance robustness, multi-technology \ac{CACC} systems integrate heterogeneous \ac{V2X} technologies, including \ac{DSRC}, \ac{C-V2X}, \ac{VLC}, and mmWave~\cite{%
ishihara2015improving,%
schettler2019deeply,%
hardes2019communication,%
%
sybis2018dynamic,%
bazzi2020wireless,%
segata2021critical,%
%
coll-perales2019sub-6ghz,%
amjad2020inband,%
}.
Additional resilience can be achieved through
time-delay compensation~\cite{santini2017consensusbased},
packet loss mitigation~\cite{giordano2019joint},
predictive filtering~\cite{wu2019cooperative},
or by utilizing data from the vehicle closest to the intended source~\cite{gorospe2025toward}.
%
Moreover, fallback mechanisms are also critical for maintaining safety during degraded operation.
Typical strategies include switching to \ac{ACC} or dissolving the platoon.
As abrupt fallback may reduce safety in dense formations~\cite{tu2019longitudinal,qin2019rearend}, graceful degradation schemes have been proposed~\cite{ploeg2015graceful,wu2019cooperative}.
More advanced controllers dynamically adapt control laws (e.g., switching from constant spacing~\cite{rajamani2000demonstration} to time-based headway~\cite{ploeg2011design}) based on current link conditions~\cite{segata2023multi-technology} and even take human-driver interactions into account~\cite{kennedy2023centralized}.

In case of sudden traffic disruptions (e.g., obstacles or accidents), emergency braking protocols (e.g.,~\cite{segata2013automatic}) remain effective under \ac{CACC}~\cite{zheng2014study,jornod2018sidelink,segata2015jerk}.
It is important to note that while individual vehicle failures or sudden, driver-initiated maneuvers within a platoon may pose risks, such hazards are not unique to platooning and are equally present in conventional traffic scenarios.
Should a platoon dissolve, our assignment algorithm can be re-applied to form new platoons once the situation stabilizes.

\subsection{Misestimation of Platooning Benefits \& Trip Costs}%
\label{sec_discussion_misestimation}

\noindent 
Our \tripcostbased{} approach uses the desired driving speed for estimating the remaining trip cost during the decision-making.
The actual driving speed of vehicles can be (much) lower than the desired driving speed due to traffic effects.
This will cause misestimations of the individual trip cost, the maneuver cost, and platoon benefits (mainly in travel time savings), thereby also of the remaining trip cost.
Additionally, updates to a platoon's desired speed upon joins and leaves, if considered, will lead to a deviation from the speed used during estimation.
In fact, \similaritybased{} suffers from the same effects after decision-making.
While we did observe small misestimations of join maneuver costs (roughly \SI{1}{\percent}), maneuvers make up for only a small part of the \SI{50}{\km} trips, and can thus be neglected.
%
A potential solution to reduce misestimation of the actual driving speed is to apply a correction coefficient to the desired driving speed of platooning opportunities.
This coefficient should depend on the current traffic situation (e.g., density or flow) and, in reality, could be estimated directly by vehicles.
%
While this correction coefficient could reduce misestimations, it makes vehicles more conservative in joining platoons at the same time.
%

Additionally, misestimation of trip costs can occur due to underestimating the distance vehicles travel in a platoon.
Vehicles calculate the distance shared with the platoon before it fully splits due to other members leaving.
If other vehicles with a destination further than the one of the ego vehicle join the same platoon, the platoon will not automatically split fully at the initially estimated location, as it still has other members.
Thus, the vehicle will stay longer in the platoon than estimated (i.e., until it reaches its destination).
This situation generally results in greater platoon benefits, as the desired driving speed of a platoon remains unchanged when new vehicles join, and a longer distance is traveled with reduced air drag.


\subsection{Time Monetization}%
\label{sec_discussion_time_monetization}

\noindent 
We base our time monetization on income as a proxy for productivity, which correlates with individuals' pace of life and travel behavior.
To account for drivers' heterogeneity, we assign monetary values to travel time using a distribution rather than a fixed, arbitrary value.
In reality, individuals would choose such monetary value, but may overestimate their opportunity costs.
One opportunity to avoid that would be to insert an hourly wage based on the payslip as an upper boundary for the cost of time for each individual.
%
While we use real-world income data to model opportunity cost, validating the correlation between this cost and actual driving behavior with real-world traffic data remains challenging.
This difficulty arises because individual driving behaviors are significantly influenced by platooning and related maneuvers, which create synthetic traffic conditions that differ from those captured in conventional traffic datasets~\cite{sommer2008progressing}.

Our approach abstracts from personal preferences beyond opportunity costs.
For example, some drivers may enjoy speeding despite low opportunity costs, while others may prioritize fuel savings even when their time is highly valued.
%
Additional factors -- such as the total amount of spare time or the perceived joy of ``leisure time'' -- can also influence how individuals value time.
Moreover, drivers may respond differently to peer effects (e.g., seeing many vehicles in a platoon)~\cite{hui1991perceived} or to moral considerations, such as guilt over not minimizing energy consumption and pollution.

While heterogeneity in individual preferences certainly exists~\cite{hoffmann2022seizing}, these preferences are inherently subjective and typically known only to the individual.
As a result, we do not explicitly model strong subjective aspects such as driving comfort.
Although proxies like driving smoothness or time spent as a follower in a platoon are measurable and likely correlate with comfort~\cite{sturm2020taxonomy}, they are not necessarily the most relevant factors in the context of automated driving systems~\cite{sturm2020taxonomy}.
Moreover, their value is again dependent on individual perception -- some drivers may regard following time as restful ``leisure time'', while others may view it as a loss of autonomy.
In both cases, we currently lack a meaningful way to assign personal valuation to these proxies.
This limitation is not unique to our monetization-based approach, but applies broadly to attempts to incorporate comfort as an optimization criterion.
Doing so would require either additional assumptions not grounded in empirical evidence or actual empirical data on subjective valuation, which falls outside the scope of this work.

In addition to inter-individual differences, situational factors -- such as
current conditions (e.g., road, weather, traffic),
trip purpose (e.g., business or leisure),
and driver mood (e.g., relaxed or agitated)
--
can further influence drivers' value of time and, consequently, their desired driving speed.
Previous empirical findings in this area, however, remain inconclusive.
For instance, \textcite{tseng2013speeding} finds more speeding for business-related trips, whereas \textcite{zhang2013risk,zhang2014traffic} find more speeding on weekends than on workdays.

Given the ambiguities and the lack of reliable, individual-level data on the influence of such factors on time valuation, we concentrate on income -- a variable with a well-established correlation to desired speed.
Nonetheless, real-world preferences vary across drivers irrespective of income and may also fluctuate within individuals based on context.
Many of the personal preferences, however, are at least indirectly captured through the concept of opportunity cost, which remains central to our modeling approach.

The decision to use systems like \ac{ACC} and platooning also depends on psychological aspects such as perceived safety \& comfort, perceived control, and mental effort~\cite{%
vasile2023comfort,%
ajzen1991theory,%
hui1991perceived,%
mick1998paradoxes,%
} that affect drivers' acceptance of the technologies.
These aspects are not captured in our approach due to the challenge of quantifying subjective trust or discomfort.
Nevertheless, future extensions could include psychological preferences as constraints (e.g., enforcing a minimum accepted gap or maximum accepted deceleration) or through probabilistic adoption models reflecting heterogeneous acceptance.

In this work, we generally assume a \SI{100}{\percent} penetration rate and that drivers will adopt platooning if it yields net utility (lower cost), based on their personal preferences.
While more detailed behavioral models could enhance realism, they would complement rather than contradict our current approach.
As shown in \cref{sec_evaluation}, our proposed platoon formation algorithm functions even in high density traffic and extreme time cost distributions.

\subsection{Application to Electric Vehicles}%
\label{sec_discussion_electric_vehicles}

\noindent 
To apply our trip cost metric and platoon formation algorithm to \acp{EV}, the fuel component must be adjusted.
Specifically, fuel consumption $\text{fuel}_{\text{trip}}$ and fuel price $C_{\text{fuel}}$ can be replaced with energy consumption and recharging costs.
Additionally, the time spent recharging and waiting at charging stations can be incorporated into the overall travel time (cf.\ \cite{schoenberg2019planning}).
An important aspect that requires further investigation is that electric vehicles can recuperate energy during braking, which can affect both the final trip cost and platooning decisions due to the potential overestimation of energy consumption.
As frequent acceleration \& deceleration maneuvers caused by vehicle interactions can lead to more energy recuperation~\cite{koch2021accurate}, automated driving using \ac{ACC} in crowded conditions (i.e., mid to high traffic densities) might become more advantageous.
However, the overall benefit will still depend on the driver's time cost value.
The impact of energy consumption on the trip cost metric is significant only when time costs are low to moderate.
Overall, we would expect reduced gains from platooning in scenarios with crowded traffic, but significant benefits in other cases (e.g., very low and very high traffic densities, or high time cost values).

%

\section{Conclusion}%
\label{sec_conclusion}

\noindent 
The benefits of \ac{ACC} and platooning have been researched for decades.
While classic \ac{ACC} became the \emph{de facto} standard for all new cars as well as for (semi-)automated driving on the freeway, debate about the lasting impact of platooning is still ongoing.
Both offer macroscopic and microscopic benefits such as improved traffic flow and safety as well as reduced fuel consumption.
While these benefits are rather well understood for trucks and utilized in truck platooning, they are less clear for passenger cars.
Accordingly, it remains unclear how to form platoons of passenger cars that optimize the personal benefit for the individual driver.

To this end, in this paper, we proposed a novel platoon formation algorithm that optimizes the personal benefit for drivers of individual passenger cars.
For computing \vtp{} assignments, the algorithm utilizes a new metric that we propose to evaluate the personal benefits of various driving systems, including platooning.
By combining fuel and travel time costs into a single monetary value, drivers can estimate overall trip costs according to a personal monetary value for time spent.
This provides an intuitive way for drivers to understand and compare the benefits of driving systems like human driving, \ac{ACC}, and, of course, platooning.
To the best of our knowledge, this is the first study to evaluate and optimize drivers' personal benefits from driving systems using a dedicated metric and an algorithm for platoon formation.

Our \vtp{} assignment algorithm estimates trip costs for both individual driving with \ac{ACC} and platooning, selecting the option with the lowest cost.
It considers both join maneuver costs and the platoon distance for more accurate estimations.
Thereby, it optimizes individual drivers' personal benefits based on their personal value for time spent.
%
Results show that while \ac{ACC} consistently outperforms human driving, it lacks behind platooning, especially in medium to high-density traffic.
Our algorithm for platoon formation outperforms traditional similarity-based platooning by as much as \SI{6}{\percent}, even in high density traffic and extreme time cost distributions.
It optimizes drivers' personal benefits by balancing fuel savings and faster travel times, particularly well when many drivers have similar time costs.
Unlike traditional similarity-based approaches, our algorithm is more flexible by performing platoon formation only when beneficial, and without strict constraints for filtering opportunities.


In future work, we want to investigate how we can avoid misestimations of platooning benefits and trip costs by detecting mismatches in speed caused by traffic effects and leaving the platoon at the initially planned location.
We further aim to allow the algorithm to continuously estimate the cost for the remaining trip in order to react to changes in platoon benefits.
Alongside these optimizations, we have yet to understand the impact of the total trip length on the platooning benefits.
To better capture real-world dynamics, we plan to adopt more sophisticated traffic simulation models that incorporate driver heterogeneity and stochastic variability.
Finally, we plan to consider more aspects of time monetization, such as
personal preferences 
and time-invariant as well as time-variant aspects. 

%

\section{Acknowledgement}%
\label{sec_acks}

\noindent 
The authors would like to thank Adam Wolisz for his valuable feedback during the preparation of this paper.

%

\printbibliography{}

%

\begin{IEEEbiography}[{\includegraphics[width=1in,height=1.25in,clip,keepaspectratio]{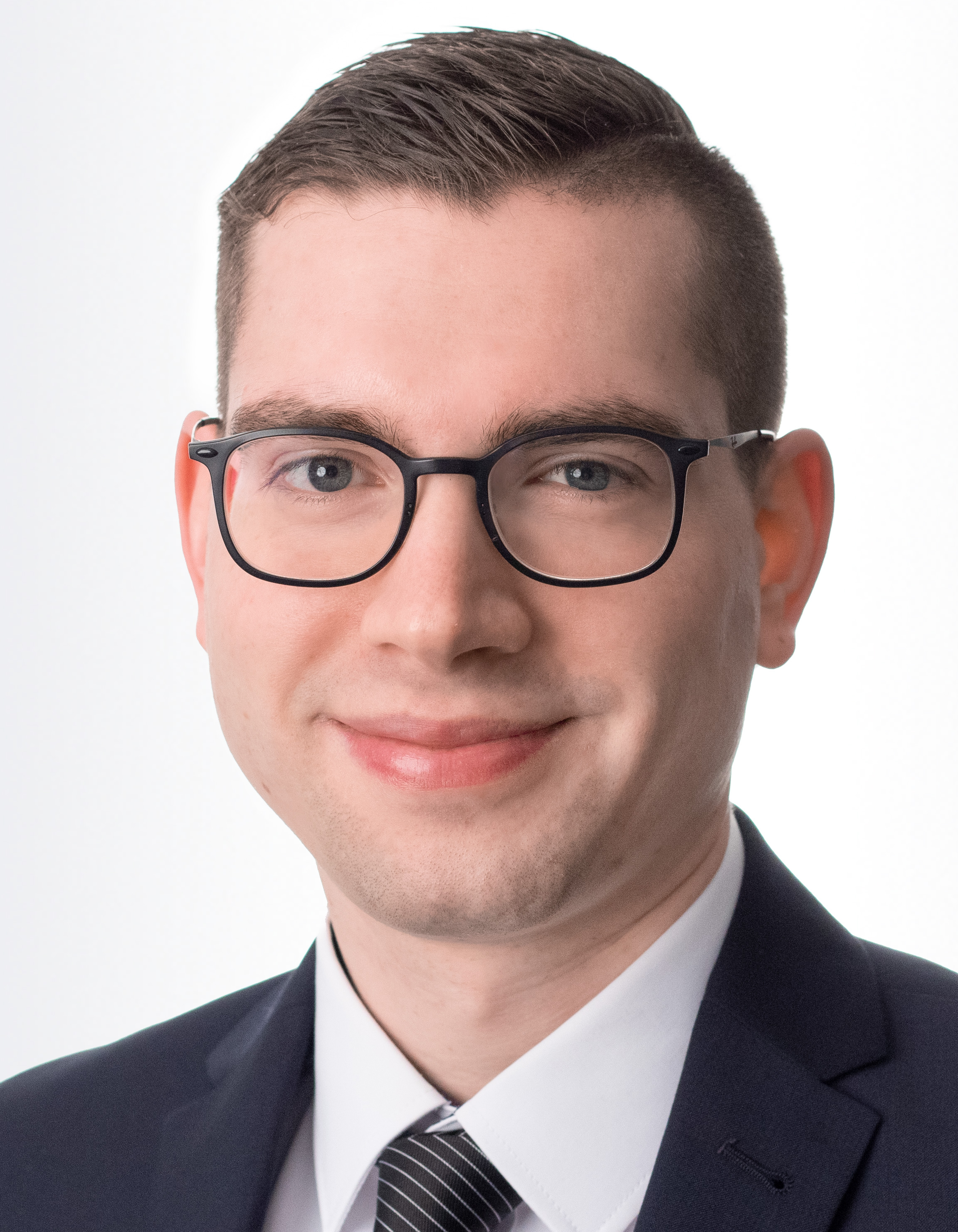}}]{Julian Heinovski}~%
(Graduate Student Member, IEEE)
is a PhD candidate and researcher in the Telecommunications Networks (TKN) group at the School of Electrical Engineering and Computer Science, TU Berlin, Germany.
He received his B.Sc.\ and M.Sc.\ degrees from the Dept.\ of Computer Science, Paderborn University, Germany, in 2016 and 2018, respectively.
Julian is an IEEE Graduate Student Member and an ACM Student Member as well as a Member of IEEE Intelligent Transportation Systems Society (ITSS) and IEEE Vehicular Technology Society (VTS).
He served as a reviewer for various manuscripts in the field of vehicular networks, cooperative driving, and intelligent transportation systems.
His research interest is in cooperative driving and intelligent transportation systems, with a focus on vehicular platooning.
\end{IEEEbiography}

\begin{IEEEbiography}[{\includegraphics[width=1in,height=1.25in,clip,keepaspectratio]{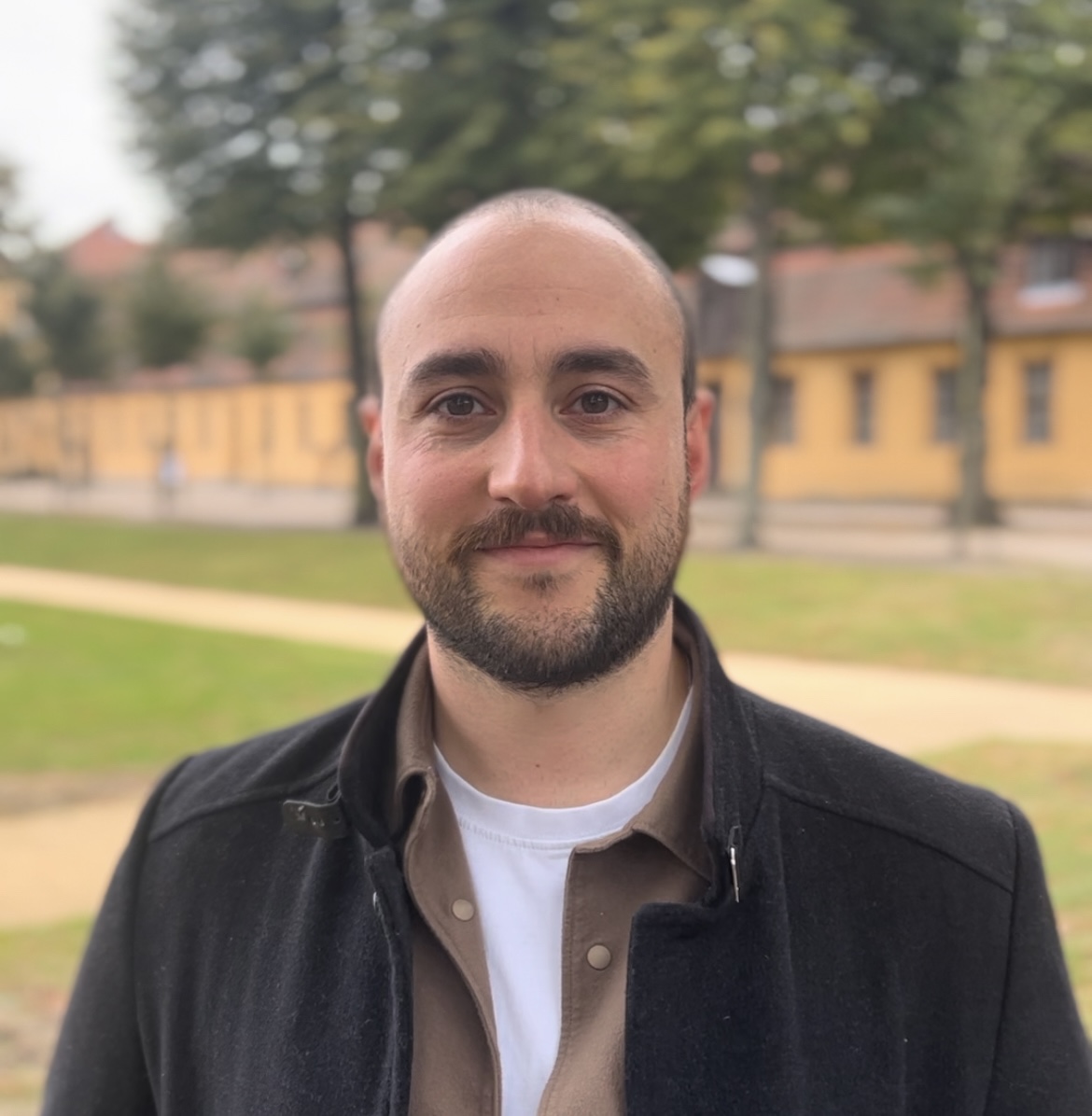}}]{Do\u{g}analp Ergen\c{c}}~%
(Member, IEEE)
is a postdoctoral researcher at Technische Universität Berlin in the Telecommunication Networks Group (TKN).
He completed his PhD in 2023 at Universität Hamburg, where he specialized in IEEE 802.1 Time-Sensitive Networking standards.
He has worked on several funded research projects in Germany and Turkey and regularly serves as a technical committee member and organizer for conferences such as IEEE RNDM, ICCN, and ICNP.
His current research centers on resilient and time-sensitive wireless networks.
\end{IEEEbiography}

\begin{IEEEbiography}[{\includegraphics[width=1in,height=1.25in,clip,keepaspectratio]{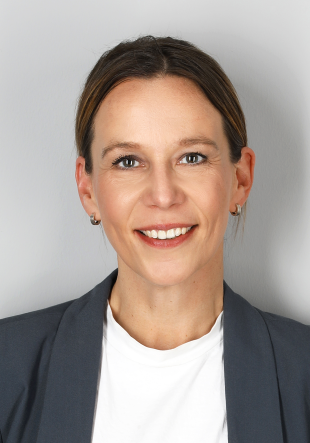}}]{Kirsten Thommes}~%
is a full professor of Organizational Behavior at Paderborn University, Germany.
She received her Diploma in 2003 (Philipps-University Marburg) and her PhD in 2008 from the Friedrich-Schiller University in Jena.
Her research interests are human-machine interaction and, among others, designing and explaining technical assistants to support humans in saving fuel consumption.
\end{IEEEbiography}

\begin{IEEEbiography}[{\includegraphics[width=1in,height=1.25in,clip,keepaspectratio]{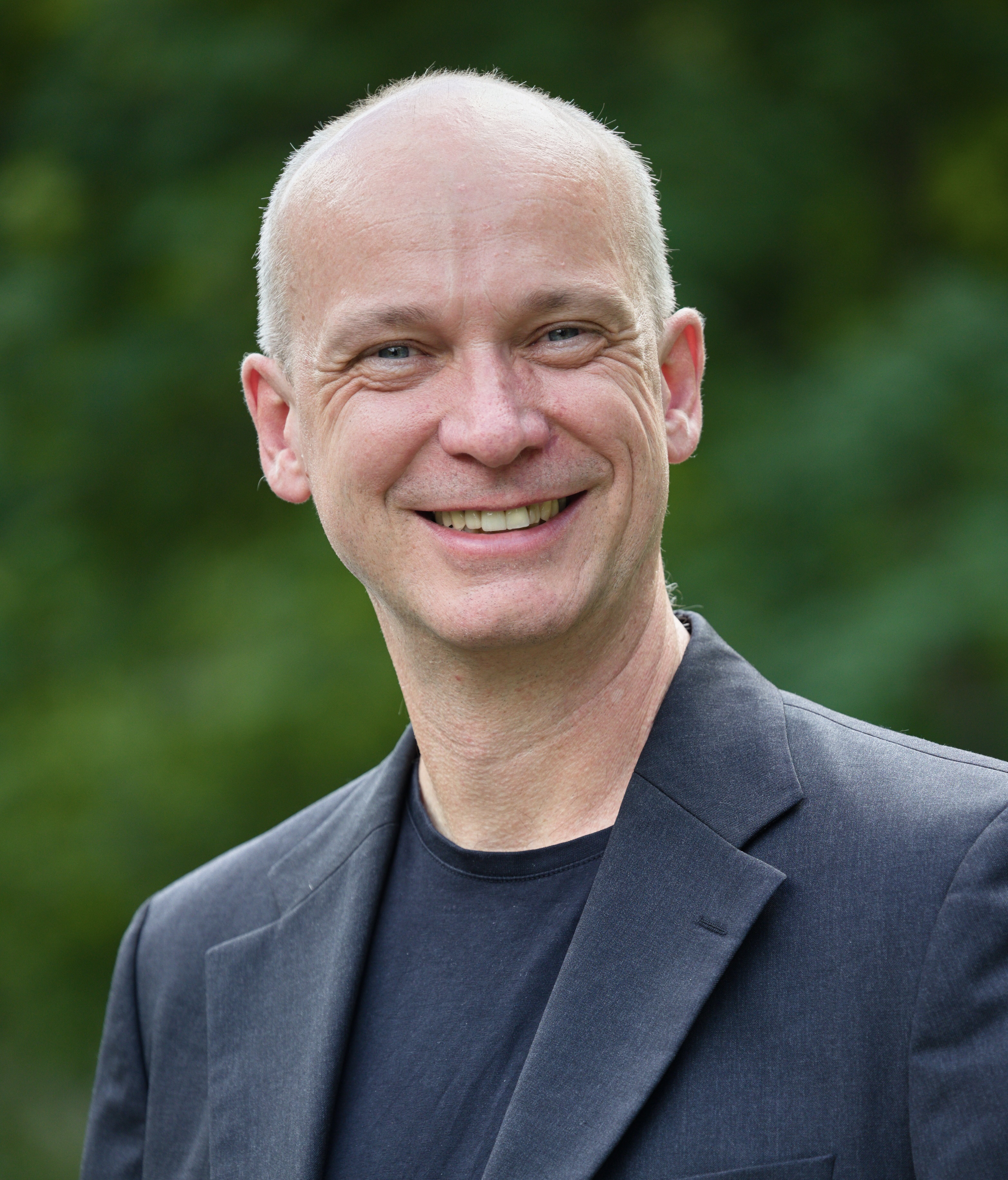}}]{Falko Dressler}~%
(Fellow, IEEE)
is full professor and Chair for Telecommunication Networks at the School of Electrical Engineering and Computer Science, TU Berlin.
He received his M.Sc.\ and Ph.D.\ degrees from the Dept.\ of Computer Science, University of Erlangen in 1998 and 2003, respectively.
Dr.\ Dressler has been associate editor-in-chief for IEEE Trans.\ on Mobile Computing and Elsevier Computer Communications as well as an editor for journals such as IEEE/ACM Trans.\ on Networking, IEEE Trans.\ on Network Science and Engineering, Elsevier Ad Hoc Networks, and Elsevier Nano Communication Networks.
He has been chairing conferences such as IEEE INFOCOM, ACM MobiSys, ACM MobiHoc, IEEE VNC, IEEE GLOBECOM.
He authored the textbooks Self-Organization in Sensor and Actor Networks published by Wiley \& Sons and Vehicular Networking published by Cambridge University Press.
He has been an IEEE Distinguished Lecturer as well as an ACM Distinguished Speaker.
Dr.\ Dressler is an IEEE Fellow as well as an ACM Distinguished Member.
He is a member of the German National Academy of Science and Engineering (acatech).
He has been serving on the IEEE COMSOC Conference Council and the ACM SIGMOBILE Executive Committee.
His research objectives include adaptive wireless networking (sub-6GHz, mmWave, visible light, molecular communication) and wireless-based sensing with applications in ad hoc and sensor networks, the Internet of Things, and Cyber-Physical Systems.
\end{IEEEbiography}

\end{document}